\documentclass[12pt]{article}
\usepackage{a4wide}
\usepackage{latexsym}
\usepackage{amsmath}
\usepackage{amsfonts}
\usepackage{amscd}
\usepackage{amssymb}
\usepackage{cite}
\usepackage{placeins}

\usepackage{pslatex}
\usepackage{graphicx}
\usepackage[latin1,utf8]{inputenc}
\usepackage[T1]{fontenc}

\allowdisplaybreaks

\newcommand{\bq}{\begin{eqnarray}}
\newcommand{\eq}{\end{eqnarray}}

\newcommand{\NF}{N_F}
\newcommand{\NO}{N_O}

\begin{document}

\thispagestyle{empty}

\begin{flushright}
  MITP/23-020
\end{flushright}

\vspace{1.5cm}

\begin{center}
  {\Large\bf Real time lattice correlation functions from differential equations \\
  }
  \vspace{1cm}
  {\large Federico Gasparotto, Stefan Weinzierl and Xiaofeng Xu \\
  \vspace{1cm}
      {\small \em PRISMA Cluster of Excellence, Institut f{\"u}r Physik, }\\
      {\small \em Johannes Gutenberg-Universit{\"a}t Mainz,}\\
      {\small \em D - 55099 Mainz, Germany}\\
  } 
\end{center}

\vspace{2cm}

\begin{abstract}\noindent
  {
We report on an exact calculation of lattice correlation functions on a finite four-dimensional lattice with 
either Euclidean or Minkowskian signature.
The lattice correlation functions are calculated by the method of differential equations.
This method can be used for Euclidean and Minkowskian signature alike.
The lattice correlation functions have a power series expansion in $1/\sqrt{\lambda}$, 
where $\lambda$ is the coupling. We show that this series is convergent for all non-zero values of $\lambda$.
At small coupling we quantify the accuracy of perturbative approximations.
At the technical level we systematically investigate the interplay between twisted cohomology 
and the symmetries of the twist function.
   }
\end{abstract}

\vspace*{\fill}

\newpage

\section{Introduction}
\label{sect:introduction}

Standard lattice calculations are usually performed by Monte Carlo methods in Euclidean signature.
In Euclidean signature the action is real and $\exp(-S_E)$ provides a factor with exponential fall-off 
and no oscillations.
The situation is more complicated if we analytically continue back the action $S_E$ to Minkowskian signature
to obtain the Minkowskian action $S_M$. 
The Minkowskian action $S_M$ is purely imaginary and $\exp(-S_M)$ provides an oscillating factor.
Evaluating integrals with oscillating integrands by Monte Carlo methods is difficult. 

In this paper we calculate lattice integrals not by Monte Carlo techniques, but by the method of differential
equations \cite{Gasparotto:2022mmp}.
We exploit the fact, that due to integration-by-parts identities 
the set of all lattice integrals for a given finite lattice spans a finite dimensional 
vector space \cite{Weinzierl:2020nhw}. 
We first derive a system of differential equations with respect to one parameter of the action 
and then solve this first-order system of differential equations with appropriate boundary conditions.
This method has the advantage that it can be applied to Euclidean or Minkowskian signature alike.
In particular, it gives us a method to compute directly lattice correlations functions with Minkowskian signature.
The method circumvents in a fundamental way the sign problem of lattice Monte Carlo simulations.
The mathematical framework underlying integration-by-parts identities is twisted cohomology \cite{aomoto1975,Matsumoto:1994,cho1995,matsumoto1998,Ohara:2003,Goto:2013,Goto:2015aaa,Goto:2015aab,Goto:2015aac,Matsumoto:2018,Matsubara-Heo:2019,Aomoto:book,Yoshida:book,Matsubara-HeoPadova}.
In recent years, twisted cohomology was recognised as a useful tool in particle physics \cite{Mizera:2017cqs,Mizera:2017rqa,Mizera:2019gea,Mastrolia:2018uzb,Frellesvig:2019kgj,Frellesvig:2019uqt,Mizera:2019vvs,Weinzierl:2020xyy,Mizera:2019ose,Mastrolia:2022tww,Frellesvig:2020qot,Chestnov:2022alh,Chestnov:2022xsy,Chen:2020uyk,Chen:2022lzr,Caron-Huot:2021xqj,Caron-Huot:2021iev,Giroux:2022wav,Cacciatori:2021nli,Cacciatori:2022mbi,Fontana:2023amt}. 

Our standard example will be $\phi^4$-theory.
This theory has been studied in great detail in the continuum and on a lattice, 
see for example \cite{Glimm:1974ki,Glimm:1975tw,Luscher:1987ay,Luscher:1987ek,Luscher:1988uq,Kleinert:2001ax,Serone:2018gjo,Serone:2019szm}.
Let us stress that we are studying a finite lattice.
For simplicity we take $L$ lattice points in each direction.
Therefore the set of all lattice points corresponds to ${\mathbb Z}_L^D$, where ${\mathbb Z}_L=\{0,\dots,L-1\}$ and $D$ denotes the space-time dimension.
In refs.~\cite{Luscher:1987ay,Luscher:1987ek,Luscher:1988uq} the lattice ${\mathbb Z}^D$ with countable many points is considered.
While a finite lattice ${\mathbb Z}_L^D$ is of course a much coarser approximation, it corresponds to the situation studied by numerical Monte Carlo simulations.
What is important to us is the fact that the vector space of all lattice integrals is finite dimensional for a finite lattice.
This is no longer the case for a lattice with countable many points.
On a finite lattice we are able to show that the lattice integrals have a power series expansion in $1/\sqrt{\lambda}$ (where $\lambda$ denotes the coupling) with an infinite radius of convergence.

While the method of differential equations can be applied -- in theory at least -- to any finite lattice in any dimension,
it should be mentioned that the method has a practical drawback: 
It is based on the finite-dimensional vector space of lattice integrands modulo integration-by-parts identities.
Although this vector space is finite dimensional, the dimension of this vector space grows exponentially with the number of lattice
points.
To give an example, the dimension of this vector space is given for $\phi^4$-theory on a lattice with $N$ points by
$3^N$.
This limits the applicability to small lattices in low dimensions.

However, we are not really interested in the vector space of lattice integrands 
but in vector space of lattice integrals.
There can be distinct lattice integrands leading to the same lattice integrals.
This originates from symmetries of the lattice action.
In this paper we systematically investigate the interplay between twisted cohomology 
and the symmetries of the twist function.
It turns out that we may partition the basis of the vector space of lattice integrands into orbits under the action
of the symmetry group and that only the number of orbits determines the size of our system.
This is a much smaller number.

The core of our method is the differential equation for the lattice integrals with respect to one (or more) parameters of the action.
It is advantageous to set up the framework such that
(i) the boundary values for the solution of the differential equation can be obtained easily and
(ii) singularities along the integration path are avoided.
This can be achieved by introducing an auxiliary flow parameter $t$ in the action, such that $t=1$ corresponds to the action
of interest and $t=0$ corresponds to a simple action, whose correlation functions can be computed easily.
This idea is inspired by the idea of ``auxiliary mass flow'' 
used in the context of Feynman diagram computations \cite{Papadopoulos:2014lla,Liu:2017jxz,Liu:2022mfb,Liu:2022chg}.
We derive a differential equation with respect to the auxiliary flow parameter $t$.
For the solution we integrate the differential equation from $t=0$ to $t=1$. 
It turns out that this differential equation has only a singularity at $t=\infty$ and 
therefore the lattice integrals have a convergent power series
expansion for all $t \in {\mathbb C}$.
Each power of the auxiliary flow parameter $t$ is always accompanied by $1/\sqrt{\lambda}$.
Setting $t=1$ gives the convergent power series
expansion in $1/\sqrt{\lambda}$.

This does not imply that it is a fast converging series.
The situation can be compared to the power series expansion of the function $\exp(-c/\sqrt{\lambda})$ in $1/\sqrt{\lambda}$:
Also this function has a convergent power series expansion in $1/\sqrt{\lambda}$ for all non-zero values of $\lambda$.
However, for small values of $\lambda$ we need to sum up sufficient many terms (with significant cancellations among them) 
before the factorial growth in the denominator outweighs the exponential growth of the numerator.
We have some freedom in setting up the action with the auxiliary flow parameter.
We find that a small modification (i.e. not including the $\phi^2$-term in the flow term)
significantly improves the convergence.

With these techniques we perform the calculation of lattice correlation functions for $\phi^4$-theory in
four space-time dimensions with either Euclidean or Minkowskian signature for a lattice with $N=16$ points (two points in each direction).
The required computing resources are moderate: 
We carried out this calculation on a single desktop with $16 \; \mathrm{GB}$ RAM within a few days.
The system of differential equations is given by a matrix of size $66524 \times 66524$. 
This matrix is sparse and fits into the memory.
We also present the corresponding results in one, two and three space-time dimension.
The associated systems of differential equations are of size $4 \times 4$, $13 \times 13$ and $147 \times 147$, 
respectively.
From a computational perspective the analytic calculation of lattice integrals 
for one, two or three space-time dimensions is cheaper.

With the analytical results at hand we may at small coupling quantify the accuracy of perturbative approximations.
In the Euclidean case the essential features of lattice integrals as a function of the coupling $\lambda$ 
can be discussed by considering the function
\bq
 A + B e^{-\frac{c}{\lambda}},
 \;\;\;\;\;\; c > 0,
\eq
where $A$ (and $B$) are slowly varying functions of $\lambda$. 
For a qualitative discussion we may treat them as constants.
At small coupling, the second term is exponentially suppressed and perturbation theory accurately predicts $A$.
The situation is different in the Minkowskian case.
Here, the lattice integrals are as a function of the coupling $\lambda$ similar to
\bq
 A + B e^{i \frac{c}{\lambda}},
 \;\;\;\;\;\; c > 0.
\eq
The difference is the additional factor $i$ in the exponent.
The second term is now oscillatory.
Perturbation theory again accurately predicts $A$, but gives no information on the second term.
The accuracy of a perturbative calculation is therefore limited by the amplitude $B$ of the oscillations.
Our method gives the analytic results and we may therefore exactly quantify the accuracy of perturbative approximations.

This paper is organised as follows:
In section~\ref{sect:notation} we set up our notation.
Twisted cohomology is introduced in section~\ref{sect:twisted_cohomology}.
Symmetries are discussed in section~\ref{sect:symmetries}.
The analytic calculation is discussed in section~\ref{sect:analytic_calculation}.
Numerical results are presented in section~\ref{sect:numerical_results}.
Finally, our conclusions are given in section~\ref{sect:conclusions}.
In appendix~\ref{sect:efficiency} we discuss the efficient reduction to master integrands.


\section{Setup and notation}
\label{sect:notation}

We are interested in the lattice discretised version of a scalar $\phi^4$-theory with Lagrange density
\bq
 {\mathcal L} & = & 
 \frac{1}{2} \left( \partial_\mu \tilde{\phi} \right) \left( \partial^\mu \tilde{\phi} \right) 
 - \frac{1}{2} \tilde{m}^2 \tilde{\phi}^2
 - \tilde{\lambda} \tilde{\phi}^4
\eq
on a space with Lorentzian signature $(1,D-1)$.
Throughout this paper we will use the terms ``Lorentzian signature'' and ``Minkowskian signature'' as synonyms.
We use a tilde to denote quantities in the continuum.
The associated Euclidean theory is obtained by a Wick rotation.
In this paper we consider an arbitrary Wick rotation angle $\delta$, such that $\delta \rightarrow 0$ corresponds to Lorentzian signature
and $\delta=\frac{\pi}{2}$ to Euclidean signature.
\begin{figure}
\begin{center}
\includegraphics[scale=1.0]{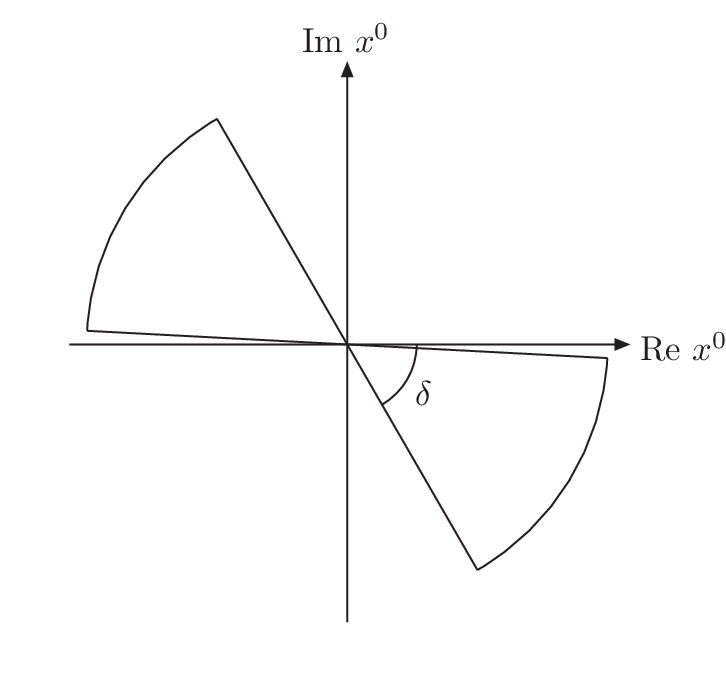}
\end{center}
\caption{
Wick rotation by an angle $\delta$ in position space.
Causality dictates that the Lorentzian contour is displaced by an infinitesimal angle. 
}
\label{fig_Wick_rotation}
\end{figure}
Causality or Feynman's $i0$-prescription dictates that the Lorentzian contour is displaced by an infinitesimal angle.
Quantities on a space with Lorentzian signature are obtained from the limit $\delta \rightarrow 0$.
The Wick rotation in position space is shown in fig.~(\ref{fig_Wick_rotation}).
It will be convenient to introduce the complex parameter
\bq
 \alpha & = & e^{i\delta}.
\eq
The complex phase $\alpha$ parametrises the Wick rotation angle $\delta$.

Let us now turn to the lattice formulation.
We consider a lattice $\Lambda$ with lattice spacing $a$ in $D \in {\mathbb N}$ space-time dimensions.
For simplicity we assume that the lattice consists of $L$ points in any direction. 
We assume periodic boundary conditions.
The lattice has $N=L^D$ points.
We label the lattice points by $x_1,\dots,x_N$ and 
denote the field at a lattice point $x$ by $\phi_x$.
The field at the next lattice point in the (positive) $j$-direction modulo $L$ is denoted by
$\phi_{x+a b_j}$.
The index $j$ takes values in $j \in \{0,1,\dots,D-1\}$ 
and $b_j$ denotes the unit vector in the $j$-th direction.
We consider a scalar $\phi^4$-theory with lattice action $S$ given by
\bq
\label{def_action}
 S
 & = &
 t S^{\mathrm{bilinear}} + S^{(4)}.
\eq
We introduced an auxiliary flow parameter $t$, such that $t=1$ corresponds to the standard action and
$t=0$ to an action consisting solely of the interaction term.
We write the bilinear term as
\bq
 S^{\mathrm{bilinear}}
 & = &
 S^{\mathrm{next} \; \mathrm{neighbours}} + S^{(2)},
\eq
where $S^{\mathrm{next} \; \mathrm{neighbours}}$ couples next neighbours and $S^{(2)}$ is proportional to the sum of the squares
of the field variables:
\bq
 S^{\mathrm{next} \; \mathrm{neighbours}}
 & = &
 \frac{i}{\alpha}
 \sum\limits_{x \in \Lambda}
 \left[
 \alpha^2 \phi_x \phi_{x+ab_0}
 - \sum\limits_{j=1}^{D-1}  \phi_x \phi_{x+ab_j}
 \right],
 \nonumber \\
 S^{(2)}
 & = &
 \frac{i}{\alpha}
 \left( D + \frac{m^2}{2} -1 - \alpha^2 \right) 
 \sum\limits_{x \in \Lambda}
 \phi_x^2.
\eq
The interaction term is given by
\bq
\label{interaction_term}
 S^{(4)}
 & = &
 \frac{i \lambda}{\alpha}
 \sum\limits_{x \in \Lambda}
 \phi_x^4.
\eq
The parameter $\alpha$ (or equivalently the angle $\delta$) parametrises the angle of the Wick rotation: 
The values $\alpha=i$ and $\delta=\frac{\pi}{2}$ correspond to an Euclidean action,
in which case the action reduces to
\bq
\label{def_Euclidean_action}
 S_E
 & = &
 \sum\limits_{x \in \Lambda}
 \left[
 - t \sum\limits_{j=0}^{D-1}  \phi_x \phi_{x+ab_j}
 + t \left( D + \frac{m^2}{2} \right) \phi_x^2
 + \lambda \phi_x^4
 \right].
\eq
The limit $\delta \rightarrow 0$ corresponds to the Minkowskian action
\bq
\label{def_Minkowski_action}
 S_M
 & = &
 i 
 \sum\limits_{x \in \Lambda}
 \left[
 t \phi_x \phi_{x+ab_0}
 - t \sum\limits_{j=1}^{D-1}  \phi_x \phi_{x+ab_j}
 + t \left( D + \frac{m^2}{2} -2 \right) \phi_x^2
 + \lambda \phi_x^4
 \right].
\eq
In the limit $\delta \rightarrow 0$ we have $\alpha=1$.
The Euclidean action is real, 
the Minkowskian action is purely imaginary 
and for a generic value $\delta \in ]0,\frac{\pi}{2}[$ the action is in general complex.
We are interested in the lattice integrals
\bq
\label{lattice_integral}
 I_{\nu_1 \nu_2 \dots \nu_N}\left(m^2,\lambda,\alpha,t\right)
 & = &
 \int\limits_{{\mathbb R}^N} d^N\phi \left( \prod\limits_{k=1}^N \phi_{x_k}^{\nu_k} \right) \exp\left(-S\right).
\eq
The correlation functions are then given by
\bq
 G_{\nu_1 \nu_2 \dots \nu_N}\left(m^2,\lambda,\alpha,t\right)
 & = &
 \frac{I_{\nu_1 \nu_2 \dots \nu_N}\left(m^2,\lambda,\alpha,t\right)}{I_{0 0 \dots 0}\left(m^2,\lambda,\alpha,t\right)}.
\eq 
Euclidean lattice integrals can be straightforwardly evaluated with Monte Carlo methods.
As the action is real, the factor $\exp(-S)$ provides for large fields an exponential damping. 
The situation is different for the Minkowskian action. Here, the action $S$ is purely imaginary and the factor
$\exp(-S)$ is oscillating. This makes it harder to evaluate the lattice integrals with Monte Carlo methods.

The standard Euclidean lattice integrals (i.e. the ones where the auxiliary parameter $t$ is set equal to $1$) are given by
\bq
 I_{\nu_1 \nu_2 \dots \nu_N}\left(m^2,\lambda,i,1\right),
\eq
the standard Minkowskian lattice integrals are given by
\bq
 I_{\nu_1 \nu_2 \dots \nu_N}\left(m^2,\lambda,1,1\right).
\eq
In both cases the value of the auxiliary parameter $t$ equals one.

It is possible to obtain the standard Minkowskian lattice integrals from Euclidean lattice integrals, if the latter are known for generic $t$
and if the number of lattice points in the time direction is even.
If $L$ is even, we may divide the lattice $\Lambda$ into two sublattices
\bq
 \Lambda & = & \Lambda_0 + \Lambda_1
\eq
where $\Lambda_0$ and $\Lambda_1$ are defined as follows: We write for a lattice point $x = (n_0,n_1,\dots,n_{D-1})$, 
where $n_j \in \{ 0, \dots, L-1 \}$.
Then
\bq
 \Lambda_0 & = & \left\{ \; x \in \Lambda \; | \; n_0 = 0 \bmod 2 \; \right\},
 \nonumber \\
 \Lambda_1 & = & \left\{ \; x \in \Lambda \; | \; n_0 = 1 \bmod 2 \; \right\}.
\eq
It is not too difficult to show that
\bq
\label{analytic_continuation}
 I_{\nu_1 \nu_2 \dots \nu_N}\left(m^2,\lambda,1,1\right)
 & = &
 \left( \prod\limits_{x_i \in \Lambda_1} \left(-1\right)^{\nu_i} \right)
 I_{\nu_1 \nu_2 \dots \nu_N}\left(m^2-4,i\lambda,i,i\right).
\eq
This allows us to obtain the result for the Minkowskian lattice integrals from the Euclidean lattice integrals with a general 
auxiliary parameter $t$.
The proof of eq.~(\ref{analytic_continuation}) follows from the substitution $\phi_x \rightarrow - \phi_x$ for all $x \in \Lambda_1$.


\section{Twisted cohomology}
\label{sect:twisted_cohomology}

The lattice integrals of eq.~(\ref{lattice_integral}) can be re-phrased in terms of twisted cohomology.
We define a function $u$, a one-form $\omega$
and a $N$-form $\Phi$ by
\bq
\label{def_twisted}
 u & = & \exp\left(-S\right),
 \nonumber \\
 \omega & = & d \ln u \;= \; -d S \; = \; \sum\limits_{x \in \Lambda} \omega_x d\phi_x,
 \nonumber \\
 \Phi & = & \left( \prod\limits_{k=1}^N \phi_{x_k}^{\nu_k} \right) d^N\phi.
\eq
In terms of these quantities we may rewrite the integral in eq.~(\ref{lattice_integral}) as
\bq
 I_{\nu_1 \nu_2 \dots \nu_N}
 & = &
 \int\limits_{{\mathbb R}^N} u \; \Phi.
\eq
The function $u$ defines the twist,
the one-form $\omega$ defines a covariant derivative $\nabla_\omega=d+\omega$.
We require that the function $u$ vanishes on the boundary of the integration (i.e. for $\phi_x \rightarrow \pm \infty$).
For non-vanishing $\lambda$ the behaviour on the boundary is determined by the $\phi_x^4$-term
and the condition that $u$ vanishes on the boundary translates to
\bq
\label{condition_boundary}
 \mathrm{Im}\left(e^{-i\delta} \lambda \right) & < & 0.
\eq
Two special cases for which eq.~(\ref{condition_boundary}) is fulfilled are of interest of us:
The first case is given by
\bq
\label{condition_boundary_example1}
 \delta \; \in \; \left]0,\frac{\pi}{2}\right]
 & \mbox{and} &
 \lambda \; \in \; {\mathbb R}_{>0}.
\eq
This allows us to study the lattice correlation functions as a function of the Wick rotation angle $\delta$.
The second case is given by
\bq
\label{condition_boundary_example2}
 \delta \; = \; 0
 & \mbox{and} &
 \mathrm{Im}\left(\lambda \right) \; < \; 0.
\eq
It is sufficient that we assume that $\lambda$ has an infinitesimal small negative imaginary part. 
This case has the advantage that it has one parameter less (the parameter $\delta$ is absent)
and is from a computational perspective cheaper.
However, eq.~(\ref{condition_boundary_example2}) does not include the case $\lambda=0$.
If $\lambda$ equals zero, the behaviour on the boundary is determined by the quadratic term 
and we find that
the function $u$ vanishes on the boundary provided
\bq
\label{condition_boundary_example3}
 \delta \; = \; 0
 & \mbox{and} &
 \mathrm{Im}\left(\lambda \right) \; \le \; 0
 \;\;\; \mbox{and} \;\;\;
 \mathrm{Im}\left(m^2 \right) \; < \; 0.
\eq
Let $\Xi$ be a regular differential $(N-1)$-form.
The requirement that $u$ vanishes on the boundary of the integration
leads to the integration-by-parts identities
\bq
 \int\limits_{{\mathbb R}^N} u \; \nabla_\omega \Xi & = & 0.
\eq
Phrased differently, the lattice integrals are invariant under
 \bq
\label{equivalence_relation}
 \Phi' & = & \Phi + \nabla_\omega \Xi,
\eq
for any regular $(N-1)$-form $\Xi$. In addition, $\Phi$ is $\nabla_\omega$-closed.
Therefore only the equivalence classes of $\nabla_\omega$-closed $N$-forms modulo exact ones is relevant.
We denote the equivalence class of $\Phi$ by $\langle \Phi |$.
These equivalence classes define the twisted cohomology group $H_\omega^N$.
For $\phi^4$-theory the dimension of $H_\omega^N$ is given by
\bq
 \NF & = & \dim H^N_\omega
 \; = \;
 3^N.
\eq
The dimension $\NF$ grows exponentially with the number of lattice points $N$.
For a lattice with $N=2^D$ points the corresponding numbers are shown in table~\ref{table_dim_twisted_cohomology_group}.
\begin{table}
\begin{center}
\begin{tabular}{|l|rrrr|}
\hline
 $D$ & $1$ & $2$ & $3$ & $4$ \\
\hline
 $\NF$ & $9$ & $81$ & $6561$ & $43046721$ \\
\hline
\end{tabular}
\end{center}
\caption{
The dimension of the twisted cohomology group as a function of $D$ for a lattice with two points in each direction.
}
\label{table_dim_twisted_cohomology_group}
\end{table}
A basis $\langle e_1 |, \dots, \langle e_{\NF} |$ of $H^N_\omega$ is given by \cite{Weinzierl:2020nhw}
\bq
\label{def_basis}
 \left( \prod\limits_{k=1}^N \phi_{x_k}^{\nu_k} \right) d^N\phi,
 & &
 \nu_k \; \in \;  \left\{0,1,2\right\}.
\eq
Using intersection numbers we may express any $\langle \Phi |$ as a linear combination of the basis $\langle e_i |$ \cite{Mastrolia:2018uzb}:
\bq
\label{reduction_to_basis}
 \left\langle \Phi \right| 
 & = &
 \sum\limits_{i=1}^{\NF} c_i \left\langle e_i \right|.
\eq
The coefficients $c_i$ are independent of the field variables $\phi_x$.
We denote the integration cycle by $| {\mathbb R}^N \rangle$ and refer to it as a twisted cycle.
We also write
\bq
 I_{\nu_1 \nu_2 \dots \nu_N}
 & = &
 \left\langle \Phi \left| {\mathbb R}^N \right. \right\rangle
\eq
to emphasize that the integral is a pairing between a twisted cocycle and a twisted cycle.
Thus we may write any lattice integral as
\bq
\label{reduction_to_basis_integral}
 \left\langle \Phi \left| {\mathbb R}^N \right. \right\rangle
 & = &
 \sum\limits_{i=1}^{\NF} c_i \left\langle e_i \left| {\mathbb R}^N \right. \right\rangle.
\eq
Let us denote by
\bq
\label{spanning_set}
 I_1 \; = \; \left\langle e_1 \left| {\mathbb R}^N \right. \right\rangle,
 & \ldots, &
 I_{\NF} \; = \; \left\langle e_{\NF} \left| {\mathbb R}^N \right. \right\rangle
\eq
the set of lattice integrals corresponding to the basis of twisted cocycles in eq.~(\ref{def_basis}).
This set spans the vector space of all lattice integrals. 
Let us emphasize that although $\{e_1,\dots,e_{N_F}\}$ is a basis of $H^N_\omega$, 
the set in eq.~(\ref{spanning_set}) is in general not a basis of the vector space of all lattice integrals,
as there might be linear relations due to symmetries among the lattice integrals of this set.
We will discuss examples in the next section.


\section{Symmetries}
\label{sect:symmetries}

The set of lattice integrals in eq.~(\ref{spanning_set}) spans the vector space of all lattice integrals.
Due to symmetries the dimension of this vector space can be smaller 
than the dimension of the twisted cohomology group.
A trivial example showing that there can be more relations among integrals then among integrands is the following:
\bq
 dz_1 \; \neq \; dz_2 & \mbox{but} & \int\limits_0^1 dz_1 \; = \; \int\limits_0^1 dz_2.
\eq
Exploiting the symmetries allows us to reduce the number of elements of the spanning set of lattice integrals.
In this section we investigate symmetries in detail.

Let $\vec{\phi}=(\phi_{x_1},\dots,\phi_{x_N})^T$ and $g \in \mathrm{SL}_N({\mathbb C})$.
We assume that the group element $g$ acts on $\vec{\phi}$ by matrix multiplication
\bq
 \vec{\phi}' & =& g \cdot \vec{\phi}.
\eq
Note that due to the requirement $g \in \mathrm{SL}_N({\mathbb C})$
the Jacobian of the transformation is equal to one
\bq
 \det\left(\frac{\partial \phi_{x_i}'}{\partial \phi_{x_j}} \right)
 & = &
 \det g
 \; = \; 1,
\eq
and therefore
\bq
 d^N\phi' & = & d^N\phi.
\eq
This group action induces an action on $\Phi$
\bq
\label{def_notation_action_on_Phi_I}
 \Phi'\left(\vec{\phi}'\right)
 & = &
 \Phi\left(g^{-1} \cdot \vec{\phi}' \right)
 \; = \;
 \left( \prod\limits_{j=1}^N \left(\sum\limits_{k=1}^N \left(g^{-1}\right)_{x_j x_k} \phi_{x_k}'\right)^{\nu_j} \right) d^N\phi'
\eq
and on $S$
\bq
 S'\left(\vec{\phi}'\right)
 & = &
 S\left(g^{-1} \cdot \vec{\phi}' \right).
\eq
The action is invariant under the transformation $g$ if
\bq
 S'\left(\vec{\phi}\right) & = & S\left(\vec{\phi}\right).
\eq
We are interested in the subgroup $G \subseteq \mathrm{SL}_N({\mathbb C})$, which leaves the action $S$ invariant.
It will be convenient to introduce for the induced action of eq.~(\ref{def_notation_action_on_Phi_I})
the notation
\bq
\label{def_notation_action_on_Phi_II}
 \Phi'\left(\vec{\phi}\right)
 \; = \; 
 g \cdot \Phi\left(\vec{\phi}\right) 
 \; = \;
 \Phi\left(g^{-1} \cdot \vec{\phi} \right).
\eq
Let $g \in G$ and let $\langle \Phi |$ be an integrand on which $g$ acts non-trivially:
\bq
 \left\langle \Phi'\left(\vec{\phi}\right) \right| & \neq & \left\langle \Phi\left(\vec{\phi}\right) \right|.
\eq
We then have the relation
\bq
\label{symmetry_relation}
 \left\langle \Phi' | {\mathbb R}^N \right\rangle
 & = &
 \left\langle \Phi | {\mathbb R}^N \right\rangle.
\eq
The proof is simple and follows from a change of variables and the invariance of the twist function:
\bq
 \left\langle \Phi' | {\mathbb R}^N \right\rangle
 & = &
 \int\limits_{{\mathbb R}^N} \Phi'\left(\vec{\phi}'\right) e^{-S\left(\vec{\phi}'\right)}
 \; = \;
 \int\limits_{{\mathbb R}^N} \Phi\left(g^{-1} \cdot \vec{\phi}'\right) e^{-S\left(\vec{\phi}'\right)}
 \; = \;
 \int\limits_{{\mathbb R}^N} \Phi\left(\vec{\phi}\right) e^{-S\left(g \cdot \vec{\phi}\right)}
 \nonumber \\
 & = &
 \int\limits_{{\mathbb R}^N} \Phi\left(\vec{\phi}\right) e^{-S\left(\vec{\phi}\right)}
 \; = \;
 \left\langle \Phi | {\mathbb R}^N \right\rangle.
\eq
We call eq.~(\ref{symmetry_relation}) a symmetry relation.
Symmetry relations allow us to reduce the number of elements in the spanning set for lattice integrals.
Suppose that
\bq
 I \; = \; \left\langle \Phi | {\mathbb R}^N \right\rangle
 & \mbox{and} &
 I' \; = \; \left\langle \Phi' | {\mathbb R}^N \right\rangle
\eq
are in the spanning set and $\Phi'=g \cdot \Phi$.
Then we may replace $I$ and $I'$ by
\bq
 \frac{1}{2} \left\langle \left( \Phi + \Phi' \right) | {\mathbb R}^N \right\rangle,
\eq
as
\bq
 \frac{1}{2} \left( \Phi - \Phi' \right)
\eq
integrates to zero. (Alternatively, we may only keep $\Phi$ and drop $\Phi'$.)

A second case is given as follows: Suppose $I = \langle \Phi | {\mathbb R}^N \rangle$ is in the spanning set and $\Phi'=g \cdot \Phi=c\Phi$ with $c \neq 1$.
Then $\Phi$ integrates to zero and we may eliminate $I$ from the spanning set.

Repeating this procedure we end up with a spanning set, where
\bq
 g \cdot \Phi & = & \Phi
\eq
for all $g \in G$ and all $\Phi$ from the spanning set.

Let us formalise this: Let $G' \subseteq G$ be a subgroup of $G$ and consider the spanning set of lattice integrals
as in eq.~(\ref{spanning_set}).
Note that each element $I_j$ of the spanning set is represented by $e_j$.
The group $G'$ acts on $e_j$ as in eq.~(\ref{def_notation_action_on_Phi_I}) and eq.~(\ref{def_notation_action_on_Phi_II}).
Suppose that there are $\NO$ non-zero orbits and suppose we label $e_1, e_2, \dots, e_{\NO}, \dots, e_{\NF}$ such that the first
$\NO$ elements are in distinct orbits.
For the orbits we introduce the notation
\bq
 o_j & = & \frac{1}{\left| G' \right|} \sum\limits_{g \in G'} g \cdot e_j,
 \;\;\;\;\;\;
 1 \; \le \; j \; \le \; \NO.
\eq
Then we may replace the spanning set by the smaller spanning set
\bq
 \left\{ o_1, \dots, o_{\NO} \right\}.
\eq
We call $e_1, e_2, \dots, e_{\NO}$ the seeds of the orbits $o_1, o_2, \dots, o_{\NO}$.
We have
\bq
 \left\langle o_j | {\mathbb R}^N \right\rangle
 & = &
 \left\langle e_j | {\mathbb R}^N \right\rangle.
\eq
We may write each orbit $o_j$ as
\bq
\label{orbit_explicit}
 o_j
 & = &
 \sum\limits_{k=1}^{\NF} c_{jk} e_k
\eq
and taking symmetries into account we may reduce any lattice integral to
\bq
\label{reduction_to_basis_integral_II}
 \left\langle \Phi \left| {\mathbb R}^N \right. \right\rangle
 & = &
 \sum\limits_{i=1}^{\NO} c_i \left\langle e_i \left| {\mathbb R}^N \right. \right\rangle.
\eq
Eq.~(\ref{reduction_to_basis_integral_II}) differs from eq.~(\ref{reduction_to_basis_integral})
that we only need $\NO$ elements in the spanning set instead of $\NF$ elements.
For actual calculations this is a significant gain in efficiency.

Note that we formulated this method deliberately for a subgroup $G'$ of $G$:
Our interest is to reduce the size of the spanning set to an acceptable level.
It is not necessary to find the minimal spanning set 
(which would provide a basis for the vector space of lattice integrals).
There can be situations, where a sufficiently large subgroup $G'$ can easily be found, but the determination
of the full group $G$ is hard.

We now discuss symmetries of the actions $S$, $S_E$ and $S_M$.
Apart from the global ${\mathbb Z}_2$-symmetry discussed in sub-section~\ref{sect:global_Z2_symmetry}
these symmetries are the remnants of Poincar\'e symmetry on a discrete lattice.
All symmetries discussed below leave $S^{\mathrm{next} \; \mathrm{neighbours}}$, $S^{(2)}$ and $S^{(4)}$ separately invariant,
hence they are symmetries for any value of the auxiliary flow parameter $t$.
 
\subsection{Global ${\mathbb Z}_2$-symmetry}
\label{sect:global_Z2_symmetry}

The action $S$ in eq.~(\ref{def_action}) is invariant under the global transformation
\bq
\label{def_global_symmetry}
 M & : & \phi_x' \; = \; - \phi_x
 \;\;\;\;\;\;
 \mbox{for all} \; x \; \in \; \Lambda.
\eq
For $\Phi$ of the form as in eq.~(\ref{def_basis}) we set
\bq
 \left|\nu\right| 
 & = &
 \sum\limits_{k=1}^N \nu_k.
\eq
The symmetry of eq.~(\ref{def_global_symmetry}) implies that we only have to consider $\Phi$'s with $|\nu|$ even,
the ones with odd $|\nu|$ integrate to zero.

\subsection{Translations}

Let us write $x = (n_0,n_1,\dots,n_{D-1})$ for the coordinates of a lattice point $x$.
We have $n_j \in \{0,1,\dots,L-1\}$.
A translation in the $j$-th direction by a unit lattice vector is given by
\bq
 T_j & : & n_j' = n_j + 1,
\eq
and $n_i'=n_i$ for $i \neq j$.
The new coordinates are understood modulus $L$.
This induces an action on the fields, which we again denote by $T_j$:
\bq
 T_j & : & \phi_x' 
 \; = \; 
 \phi_{T_j^{-1} x},
 \; = \; 
 \phi_{x-ab_j},
 \;\;\;\;\;\; 0 \; \le \; j \; \le \; D-1. 
\eq
The action $S$ in eq.~(\ref{def_action}) is invariant under the translations $T_j$.

\subsection{Spatial rotations}

For $D \ge 2$ we may consider rotations and for $D \ge 3$ we may consider spatial rotations.
For simplicity we always assume that the lattice consists of $L$ points in each direction.

Let $i,j \in \{1,\dots,D-1\}$ be two indices referring to two spatial directions.
Let $R_{ij}$ be the rotation, which acts on $(n_i,n_j)$ as
\bq
 \left(\begin{array}{c}
   n_i' \\
   n_j' \\
 \end{array} \right)
 & = &
 \left(\begin{array}{rr}
   0 & -1 \\
   1 & 0 \\
 \end{array} \right)
 \left(\begin{array}{c}
   n_i \\
   n_j \\
 \end{array} \right)
\eq
and trivial on all other coordinates.
The new coordinates are understood modulus $L$.
In other words, $R_{ij}$ corresponds to a rotation by an angle $\pi/2$ in the plane spanned by the basis vectors 
$b_i$ and $b_j$.
The action $S$ in eq.~(\ref{def_action}) is invariant under 
\bq
\label{def_rotation_on_phi}
 R_{ij} & : & \phi_x' \; = \; \phi_{R^{-1}_{ij} x}.
\eq

\subsection{Boosts}

For rotations in the plane spanned by one spatial direction and the time direction we have to discuss the cases
of the Euclidean actions $S_E$, the general $\delta$-dependent action $S$ and the Minkowskian action $S_M$
separately.

The Euclidean action $S_E$ in eq.~(\ref{def_Euclidean_action}) is invariant under rotations $R_{0 j}$ (as defined in eq.~(\ref{def_rotation_on_phi}))
with $j \in \{1,\dots,D-1\}$.
For the $\alpha$-dependent action $S$ of eq.~(\ref{def_action}) the $\alpha$-dependent terms spoil the invariance.
Let us now assume that $L$ is even.
Then the Minkowskian action $S_M$ of eq.~(\ref{def_Minkowski_action}) is invariant under a symmetry $R_{0 j}$, where
$R_{0 j}$ acting on $(n_0,n_j)$ is defined as above and
\bq
 R_{0j} & : & \phi_x' \; = \; 
 \left\{ \begin{array}{rl}
  \phi_{R^{-1}_{0j} x}, & \left(n_0+n_j\right) \; \mbox{even}, \\
  - \phi_{R^{-1}_{0j} x}, & \left(n_0+n_j\right) \; \mbox{odd}. \\
 \end{array} \right.
\eq

\subsection{Time reversal and spatial reversal}

Let $P_i$ be the operation, which acts on $n_i$ as
\bq
 P_i & : & n_i' \; = \; -n_i
\eq
and trivial on all other $n_j$'s.
The new coordinate $n_i'$ is understood modulo $L$.
The action $S$ in eq.~(\ref{def_action}) is invariant under 
\bq
 P_{i} & : & \phi_x' \; = \; \phi_{P^{-1}_{i} x}.
\eq

\subsection{Example}

We consider the Euclidean action $S_E$ and the Minkowskian action $S_M$ 
together with a lattice $\Lambda$ with $L=2$ points in each direction.
\begin{figure}
\begin{center}
\includegraphics[scale=0.7]{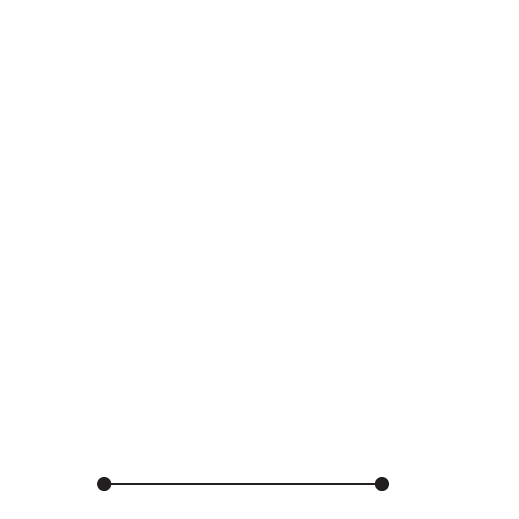}
\includegraphics[scale=0.7]{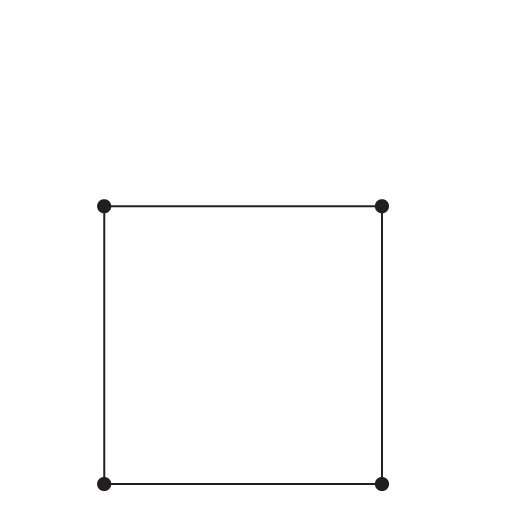}
\includegraphics[scale=0.7]{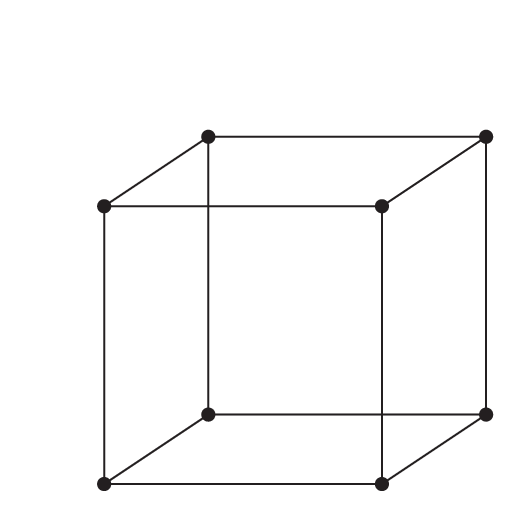}
\hspace*{5mm}
\includegraphics[scale=0.7]{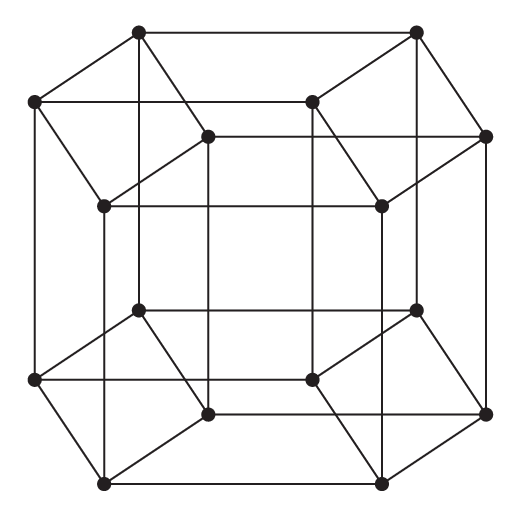}
\end{center}
\caption{
The lattices with two points in each direction for one, two, three and four space-time dimensions.
Shown are the lattice points and the links between them.
}
\label{fig_lattices}
\end{figure}
These lattices are shown for $D \in \{1,2,3,4\}$ space-time dimensions in fig.~\ref{fig_lattices}.
As group $G'$ we take the group generated by the symmetries listed above:
the global ${\mathbb Z}_2$-symmetry $M$, the translations $T_j$, the spatial rotations $R_{ij}$ 
(with $i,j$ spatial indices) and $R_{0j}$ (in the Euclidean case these are also rotations, in the Minkowskian case
these are the lattice versions of boosts).
For $L=2$ the operations $P_i$ are identical to the translations and do not give new symmetries.
The order of the group as a function of the space-time dimension $D$ is given in table~\ref{table_dim_Gprime}.
\begin{table}
\begin{center}
\begin{tabular}{|l|rrrr|}
\hline
 $D$ & $1$ & $2$ & $3$ & $4$ \\
\hline
 $|G'|$ & $4$ & $16$ & $96$ & $768$ \\
\hline
\end{tabular}
\end{center}
\caption{
The order of the symmetry group $G'$ for the Euclidean action $S_E$ or the Minkowskian action $S_M$
as a function of $D$ for a lattice with two points in each direction.
}
\label{table_dim_Gprime}
\end{table}
The order of the group $G'$ is the same in the Euclidean case as in the Minkowskian case.
\begin{table}
\begin{center}
\begin{tabular}{|l|rrrr|}
\hline
 $D$ & $1$ & $2$ & $3$ & $4$ \\
\hline
 $\NF$ & $9$ & $81$ & $6561$ & $43046721$ \\
 $\NO$ & $4$ & $13$ & $147$ & $66524$ \\
\hline
\end{tabular}
\end{center}
\caption{
The dimension of the twisted cohomology group $\NF$ 
and the number of non-zero orbits $\NO$ 
for the Euclidean action $S_E$ or the Minkowskian action $S_M$
as a function of $D$ for a lattice with two points in each direction.
}
\label{table_dim_orbits}
\end{table}
The number of non-zero orbits $\NO$ as a function of $D$ are shown in table~\ref{table_dim_orbits}.
The number of orbits is the same in the Euclidean case as in the Minkowskian case.
In this table we also repeated the dimension of the twisted cohomology group $\NF$.
We observe a significant reduction.

We recall from eq.~(\ref{orbit_explicit}) that we may write any orbit as
\bq
\label{orbit_explicit_v2}
 o_j
 & = &
 \sum\limits_{k=1}^{\NF} c_{jk} e_k.
\eq
In the Euclidean case and in the Minkowskian case we find
\bq
 \mbox{Euclidean} & : &
 c_{jk} \; \in \; \left\{ 0, \frac{\left| G_{e_j}' \right|}{\left| G' \right|} \right\},
 \nonumber \\
 \mbox{Minkowskian} & : &
 c_{jk} \; \in \; \left\{ -\frac{\left| G_{e_j}' \right|}{\left| G' \right|}, 0, \frac{\left| G_{e_j}' \right|}{\left| G' \right|} \right\},
\eq 
where $G_{e_j}'$ is the stabiliser subgroup of $e_j$.
For the Minkowskian case we have to keep track of the signs (originating from the boosts) and
we define for the decomposition as in eq.~(\ref{orbit_explicit}) or eq.~(\ref{orbit_explicit_v2})
\bq
 \mathrm{sign}\left(e_k,o_j\right)
 & = &
 \left\{ \begin{array}{ll}
  0, & \mbox{if} \; c_{jk}=0, \\
  \mathrm{sign}\left(c_{jk}\right), & \mbox{otherwise}. \\
 \end{array}
 \right.
\eq


\section{The analytic calculation}
\label{sect:analytic_calculation}

\subsection{The differential equation}
\label{sect:differential_equation}

Let us now consider the derivative of a lattice integral $I_{\nu_1 \dots \nu_N}$ 
with respect to the parameters $\{m^2,\lambda,\alpha,t\}$ of the action.
In principle we may compute the derivative with respect to any of these parameters.
It will be convenient to focus on the derivative with respect to the auxiliary flow parameter $t$.
Taking the derivative of the exponential brings down extra factors of the field variables and we obtain
for the scalar theory of eq.~(\ref{def_action})
\bq
 \frac{d}{dt} I_{\nu_1 \dots \nu_N}
 & = &
 \frac{d}{dt} \int\limits_{{\mathbb R}^N} d^N\phi \left( \prod\limits_{k=1}^N \phi_{x_k}^{\nu_k} \right) \exp\left(-S\right)
 \nonumber \\
 & = &
 - \int\limits_{{\mathbb R}^N} d^N\phi \left( \prod\limits_{k=1}^N \phi_{x_k}^{\nu_k} \right) S^{\mathrm{bilinear}}  \exp\left(-S\right).
\eq
With the help of eq.~(\ref{reduction_to_basis_integral_II})
we may re-express the right-hand side as a linear combination of the 
spanning set $\{I_1, \dots, I_{\NO}\}$.
Doing this for every element $I_i$ of the spanning set $\{I_1, \dots, I_{\NO}\}$ yields
\bq
\label{general_dgl}
 \frac{d}{dt} I_{j}
 & = &
 \sum\limits_{k=1}^{\NO}
 A_{jk} I_{k}.
\eq
This is a system of $\NO$ first-order differential equations.
Setting $\vec{I} = (I_1,\dots,I_{\NO})^T$ we write this system as
\bq
\label{general_dgl_matrix}
 \frac{d}{dt} \vec{I} & = & A \vec{I}.
\eq
In order to compute eq.~(\ref{general_dgl}) efficiently we proceed as follows:
For $1 \le j \le \NO$ we recall that $e_j$ is the seed of the orbit $o_j$.
We first need the reduction
\bq
\label{required_reduction}
 \frac{d}{dt} e_j & = & \sum\limits_{l=1}^{\NF} c_{jl} e_l + \nabla_\omega \Xi.
\eq
Efficient methods for this reduction are discussed in appendix~\ref{sect:efficiency}.
Then $A_{jk}$ in eq.~(\ref{general_dgl}) is given by
\bq
 A_{jk}
 & = &
 \sum\limits_{l=1}^{\NF} c_{jl} \cdot \mathrm{sign}\left(e_l,o_k\right).
\eq
Note that the sum in eq.~(\ref{general_dgl}) is up to $\NO$, while the sum in eq.~(\ref{required_reduction}) is up to $\NF$.


\subsection{Boundary values}
\label{sect:boundary_values}

For the analytic solution of the lattice integrals we need in addition to the differential equation eq.~(\ref{general_dgl_matrix})
boundary values at one specific point.

At the value $t_0=0$ for the auxiliary flow parameter the lattice integrals 
are independent of $m^2$ and
factorise into one-dimensional integrals
\bq
\label{boundary_integral}
 I_{\nu_1 \nu_2 \dots \nu_N}\left(m^2,\lambda,\alpha,0\right)
 & = &
 \prod\limits_{k=1}^N 
 \int\limits_{{\mathbb R}} d\phi \; \phi^{\nu_k} \exp\left(- \frac{i \lambda}{\alpha} \phi^4 \right).
\eq
We write
\bq
 I_{\nu_1 \nu_2 \dots \nu_N}\left(m^2,\lambda,\alpha,0\right)
 & = &
 \prod\limits_{k=1}^N B_{\nu_k}\left(\lambda,\alpha\right)
\eq
with
\bq
 B_{\nu}\left(\lambda,\alpha\right)
 & = &
 \left\{\begin{array}{cl}
 \frac{1}{2} \left(\frac{\alpha}{i\lambda}\right)^{\frac{\nu+1}{4}} \Gamma\left(\frac{\nu+1}{4}\right), & \nu \; \mbox{even} \\
 0, & \nu \; \mbox{odd}.
 \end{array} \right.
\eq
In particular we have that $I_{\nu_1 \nu_2 \dots \nu_N}(m^2,\lambda,\alpha,0)$ equals zero 
whenever (at least) one of the indices $\nu_j$ is odd.
 

\subsection{Convergence}
\label{sect:convergence}

The entries of the matrix $A$ in the differential equation eq.~(\ref{general_dgl_matrix}) are always polynomials
in the auxiliary flow parameter $t$.
This follows directly from the reduction algorithm: 
The reduction algorithm terminates in a finite number of steps
and each step can only introduce positive powers 
of $t$ (see eq.~(\ref{Groebner_element}) and eq.~(\ref{lower_terms})).
Hence, the entries of the matrix $A$ are holomorphic functions of $t$ in the whole complex plane ${\mathbb C}$.
It follows that the solutions $\vec{I}$ are holomorphic functions of $t$ in the whole complex plane ${\mathbb C}$
(see for example theorem 2.1 in ref.~\cite{Wasow}).
In particular, this implies that each lattice integral from the set $\{I_1,\dots,I_{\NO}\}$ has a convergent
power series expansion around $t=0$ with an infinite radius of convergence.
This series agrees with the Taylor expansion around $t=0$.
It follows from section~\ref{sect:boundary_values} that we may write $I_j \in \{I_1,\dots,I_{\NO}\}$ as
a convergent series
\bq
\label{convergent_t_series}
 I_j\left(m^2,\lambda,\alpha,t\right) & = &
 \left(\frac{1}{\sqrt{\lambda}}\right)^{\frac{\left|\nu\right|}{2}+\frac{N}{2}}
 \sum\limits_{k=0}^\infty C_{j,k}\left(m^2,\alpha\right) \cdot \left(\frac{t}{\sqrt{\lambda}}\right)^k,
\eq
where $|\nu|=\nu_1+\dots+\nu_N$. 
The coefficient functions $C_{j,k}(m^2,\alpha)$ depend only on $m^2$ and $\alpha$, but are independent of
$\lambda$.
The series in eq.~(\ref{convergent_t_series}) is convergent for all values $t \in {\mathbb C}$, and in particular for $t=1$.
Our main interest is the value at $t=1$ and we obtain the convergent series expansion
\bq
\label{convergent_lambda_series}
 I_j\left(m^2,\lambda,\alpha,1\right) & = &
 \left(\frac{1}{\sqrt{\lambda}}\right)^{\frac{\left|\nu\right|}{2}+\frac{N}{2}}
 \sum\limits_{k=0}^\infty C_{j,k}\left(m^2,\alpha\right) \cdot \left(\frac{1}{\sqrt{\lambda}}\right)^k.
\eq
We are only considering lattices with an even number of lattice points.
Furthermore, for any non-vanishing lattice integral we have that $|\nu|$ is even (if $|\nu|$ is odd
the lattice integral vanishes due to the global ${\mathbb Z}_2$-symmetry).
It follows that eq.~(\ref{convergent_lambda_series}) is a convergent series in $1/\sqrt{\lambda}$.
Any lattice integral $I_{\nu_1 \dots \nu_N}$ is a linear combination of integrals from the set $\{I_1,\dots,I_{\NO}\}$.
The coefficients in this linear combination are rational functions of $\lambda$.
From the reduction algorithm it follows that in the denominator we can only have factors of $\lambda$,
but never any non-trivial polynomial of $\lambda$.
This shows that any lattice integral $I_{\nu_1 \dots \nu_N}$ has a convergent series expansion 
in $1/\sqrt{\lambda}$ with an infinite radius of convergence.

A few remarks are in order:

\begin{enumerate}
\item The proof that the expansion in eq.~(\ref{convergent_lambda_series}) has an infinite radius of convergence 
in the variable $1/\sqrt{\lambda}$ relies on the fact that we are considering a finite lattice.
It does not carry over to a lattice ${\mathbb Z}^D$ with countable many lattice points: 
On a finite lattice we have a finite system of differential equations and the infinite radius of convergence
follows from the holomorphicity of the connection matrix $A$.
On a lattice with countable many lattice points we no longer have a finite system of differential equations.

\item The convergent series expansion is the analytic continuation of the lattice integral to regions where
the integral representation of eq.~(\ref{lattice_integral}) is not defined. 
For example, in the Euclidean case the integral representation is not defined for $\lambda < 0$, 
however the series expansion can 
be evaluated without problems for purely imaginary values of $\sqrt{\lambda}$.
The situation is similar to the definition of Euler's Gamma function, defined for 
$\mathrm{Re}\; z > 0$ by the integral representation
\bq
 \Gamma\left(z\right)
 & = & 
 \int\limits_0^\infty dt \; t^{z-1} e^{-t},
 \;\;\;\;\;\; \mathrm{Re}\; z > 0.
\eq
The Gamma function has an analytic continuation to $\mathrm{Re}\; z \le 0$.

\item In order to compute the lattice integrals we only need to determine the coefficients 
$C_{j,k}$ in eq.~(\ref{convergent_lambda_series}).
In principle this can be done as follows: We expand
\bq
 \exp\left(-t S^{\mathrm{bilinear}}\right)
 & = &
 \sum\limits_{k=0}^\infty \frac{\left(-t\right)^k}{k!} \left(S^{\mathrm{bilinear}}\right)^k.
\eq
At each order $k$ in $t$ we expand the polynomial $(S^{\mathrm{bilinear}})^k$ into a sum of monomials in the 
field variables $\phi_x$.
Each term is then of the form of a boundary integral as in eq.~(\ref{boundary_integral})
and can be evaluated as discussed in section~\ref{sect:boundary_values}.
This works for the first few terms of the expansion in eq.~(\ref{convergent_lambda_series}), but becomes soon
highly inefficient, as the computational cost increases substantially with $k$:
$S^{\mathrm{bilinear}}$ is a quadratic polynomial in $N$ field variables $\phi_x$
and expanding $(S^{\mathrm{bilinear}})^k$ into monomials will soon exceed the available memory.
To give an example: For $D=4$ and $L=2$ we have $N=16$. Expansion of $(S^{\mathrm{bilinear}})^k$ reaches its limit
for $k \gtrsim 13$ on a standard laptop.

However, there is a better way to determine the coefficients $C_{j,k}(m^2,\alpha)$: We may compute them
at constant computational cost with the help of the differential equation.
Let $d_{\mathrm{max}}$ be the maximum degree of the entries of the matrix $A$ in the variable $t$.
The differential equation leads to a recursion relation of order $(d_{\mathrm{max}}+1)$:
The coefficients $C_{j,k+1}$ can be computed from the coefficients $C_{j,k}, \dots, C_{j,k-d_{\mathrm{max}}}$.
Once the differential equation is known, this provides an efficient method to compute the coefficients $C_{j,k}$
to high order.

\item The fact that eq.~(\ref{convergent_lambda_series}) is convergent for all non-zero values of $\lambda$ does
not imply that it is for all non-zero values of $\lambda$ a fast convergent series.
This can be compared to the series expansion of
\bq
\label{example_expansion}
 \exp\left(-\frac{c}{\sqrt{\lambda}}\right)
 & = &
 \sum\limits_{k=0}^\infty \frac{1}{k!} \left(-\frac{c}{\sqrt{\lambda}}\right)^k.
\eq
Eq.~(\ref{example_expansion}) converges for all non-zero values of $\lambda$, but for $\lambda$ small
we will need many terms
before the factorial growth of $(k!)$ outweighs the exponential growth of $(-c/\sqrt{\lambda})^k$.
In addition, there might be severe cancellations between individual terms before convergence is reached.
Let us give an example: We consider the lattice integral $I_{0000}$ in two space-time dimension with Euclidean signature.
As parameters we choose $m=1$ and $\lambda=0.2$.
The value of the integral is
\bq
 I_{0000} & = & 2.10575.
\eq
We need about $560$ terms in eq.~(\ref{convergent_lambda_series}) to reach six digits accuracy. 
Individual terms are of the size up to ${\mathcal O}(10^{41})$ and we have a cancellation of about $40$ digits.
For a precision of six digits we have to compute the individual terms with at least $46$ digits.

The series in eq.~(\ref{convergent_lambda_series}) converges faster for larger values of $\lambda$.

In the next section we will discuss how to improve the convergence.

\end{enumerate}


\subsection{Efficiency improvements}
\label{sect:efficiency_improvements}

We have some freedom in setting up the action with the auxiliary flow parameter.
In eq.~(\ref{def_action}) we considered the action
\bq
 S
 & = &
 t S^{\mathrm{next} \; \mathrm{neighbours}} + t S^{(2)} + S^{(4)}.
\eq
We may consider a slight modification
\bq
 \tilde{S}
 & = &
 \tilde{t} S^{\mathrm{next} \; \mathrm{neighbours}} + S^{(2)} + S^{(4)},
\eq
where the flow parameter $\tilde{t}$ appears only as coefficient of $S^{\mathrm{next} \; \mathrm{neighbours}}$,
but not of $S^{(2)}$.
We may again derive a differential equation, this time with respect to $\tilde{t}$:
\bq
 \frac{d}{d\tilde{t}} \vec{\tilde{I}} & = & \tilde{A} \vec{\tilde{I}}.
\eq
As before, we are only interested in the case $\tilde{t}=1$ and we have 
$\vec{\tilde{I}}(m^2,\lambda,\alpha,1)=\vec{I}(m^2,\lambda,\alpha,1)$.
The entries of the matrix $\tilde{A}$ are again polynomials in $\tilde{t}$, hence we may write $\vec{\tilde{I}}$ as a
power series in $\tilde{t}$ with an infinite radius of convergence.
We write each lattice integral as
\bq
\label{series_expansion_ttilde}
 \tilde{I}_{\nu_1 \dots \nu_N}\left(m^2,\lambda,\alpha,\tilde{t}\right) & = &
 \sum\limits_{k=0}^\infty \tilde{C}_{\nu_1 \dots \nu_N, k}\left(m^2,\lambda,\alpha\right) \tilde{t}^k.
\eq
As boundary point we take $\tilde{t}=0$. As before, the lattice integrals factorise at the boundary point into
a product of one-dimensional integrals:
\bq
\label{tilde_boundary_integral}
 \tilde{I}_{\nu_1 \nu_2 \dots \nu_N}\left(m^2,\lambda,\alpha,0\right)
 = 
 \prod\limits_{k=1}^N 
 \int\limits_{{\mathbb R}} d\phi \; \phi^{\nu_k} \exp\left(- \frac{i \mu}{\alpha} \phi^2 - \frac{i \lambda}{\alpha} \phi^4 \right),
 & &
 \mu \; = \; D + \frac{m^2}{2} -1 - \alpha^2.
 \;\;
\eq
We write
\bq
\label{tilde_boundary_integral_v2}
 \tilde{I}_{\nu_1 \nu_2 \dots \nu_N}\left(m^2,\lambda,\alpha,0\right)
 & = &
 \prod\limits_{k=1}^N \tilde{B}_{\nu_k}\left(m^2,\lambda,\alpha\right)
\eq
with
\bq
 \tilde{B}_{\nu}\left(m^2,\lambda,\alpha\right)
 & = &
 \left\{\begin{array}{cl}
 \frac{1}{2} \left(\frac{\alpha}{i\lambda}\right)^{\frac{\nu+1}{4}} 
 \sum\limits_{n=0}^\infty \frac{1}{n!} \Gamma\left(\frac{n}{2}+\frac{\nu+1}{4}\right) \left(-\frac{i\mu}{\alpha} \sqrt{\frac{\alpha}{i\lambda}}\right)^n, & \nu \; \mbox{even} \\
 0, & \nu \; \mbox{odd}.
 \end{array} \right.
\eq
The sum can be evaluated in terms of modified Bessel functions of the first kind.
As before we have that $\tilde{I}_{\nu_1 \nu_2 \dots \nu_N}(m^2,\lambda,\alpha,0)$ equals zero 
whenever (at least) one of the indices $\nu_j$ is odd.
For the boundary point we therefore only need $\tilde{B}_{0}(m^2,\lambda,\alpha)$
and $\tilde{B}_{2}(m^2,\lambda,\alpha)$. These are given by
\bq
 \tilde{B}_{0}\left(m^2,\lambda,\alpha\right)
 & = &
 \frac{\pi}{2}
 \left(\frac{2 \alpha}{i\lambda}\right)^{\frac{1}{4}}
 \left(\chi^2\right)^{\frac{1}{4}}
 e^{\chi^2} 
  \left[  I_{-\frac{1}{4}}\left(\chi^2\right)
         - \frac{\chi}{\left(\chi^2\right)^{\frac{1}{2}}} I_{\frac{1}{4}}\left(\chi^2\right)
  \right],
 \\
 \tilde{B}_{2}\left(m^2,\lambda,\alpha\right)
 & = &
 \frac{\pi}{4}
 \left(\frac{2 \alpha}{i\lambda}\right)^{\frac{3}{4}}
 \left(\chi^2\right)^{\frac{3}{4}}
 e^{\chi^2} 
  \left[ I_{\frac{1}{4}}\left(\chi^2\right) + I_{-\frac{3}{4}}\left(\chi^2\right) 
         - \frac{\chi}{\left(\chi^2\right)^{\frac{1}{2}}} \left( I_{-\frac{1}{4}}\left(\chi^2\right) - I_{\frac{3}{4}}\left(\chi^2\right) \right) \right],
 \nonumber
\eq
with
\bq
 \chi & = & 
 \frac{\mu}{2} \sqrt{\frac{i}{2 \alpha\lambda}}.
\eq
$I_\nu(x)$ denotes the modified Bessel functions of the first kind.

Remark 3 from section~\ref{sect:convergence} applies also here:
In principle we may compute the coefficients $\tilde{C}_{\nu_1 \dots \nu_N, k}(m^2,\lambda,\alpha)$ 
by expanding
\bq
 \exp\left(-\tilde{t} S^{\mathrm{next} \; \mathrm{neighbours}}\right)
 & = &
 \sum\limits_{k=0}^\infty \frac{\left(-\tilde{t}\right)^k}{k!} \left(S^{\mathrm{next} \; \mathrm{neighbours}}\right)^k.
\eq
At each order $k$ in $\tilde{t}$ we then expand the polynomial $(S^{\mathrm{next} \; \mathrm{neighbours}})^k$ into a sum of monomials in the 
field variables $\phi_x$.
Each term is then of the form of a boundary integral as in eq.~(\ref{tilde_boundary_integral}) and eq.~(\ref{tilde_boundary_integral_v2}).
This can be used for the first few terms, but becomes impracticable for higher order terms in $\tilde{t}$.
As before, the solution is to use the differential equation.
The differential equation leads to a recursion relation.
With the help of this recursion relation the higher order terms in $\tilde{t}$ can be computed at constant computational cost.

We observe that any non-zero lattice integral has either an expansion in even powers of $\tilde{t}$ 
or an expansion in odd powers of $\tilde{t}$.
Hence, for a given non-zero lattice integral either all odd or all even coefficients $\tilde{C}_{\nu_1 \dots \nu_N, k}(m^2,\lambda,\alpha)$
vanish, where even/odd refers to the index $k$ being even or odd.
This pattern follows from the following three facts:
\begin{enumerate}
\item An integral of the form as in eq.~(\ref{tilde_boundary_integral}) or eq.~(\ref{tilde_boundary_integral_v2})
is only non-zero, if all indices $\nu_j$ are even.
\item The polynomial $S^{\mathrm{next} \; \mathrm{neighbours}}$ is linear in each variable $\phi_x$ and each monomial of 
the polynomial $S^{\mathrm{next} \; \mathrm{neighbours}}$
defines an edge in the lattice.
\item Every loop in the lattice has an even number of edges.
\end{enumerate}
Let's first consider a lattice integral, where all indices $\nu_j$ are even.
This integral starts at order $\tilde{t}^0$.
At order $\tilde{t}^1$ we necessarily have exactly two lattice points, where the corresponding indices are odd 
(as $S^{\mathrm{next} \; \mathrm{neighbours}}$ is homogeneous of degree $2$ and linear in each variable $\phi_x$).
These integrals vanish.
It is not too difficult to see that we only get a non-zero integral if there is a monomial obtained from expanding 
$(S^{\mathrm{next} \; \mathrm{neighbours}})^k$ which corresponds to a clothed path in the lattice through the identification of each monomial
in $S^{\mathrm{next} \; \mathrm{neighbours}}$ with an edge of the lattice.
Furthermore, one can show that any clothed path on the lattice can be constructed from two basic loops:
(i) the loop going from a lattice point to a neighbouring lattice point and back
and (ii) the basic plaquette.
The first loop consists of two edges, the second loop of four edges. It follows that the number of edges in any loop is even
and that a lattice integral, where all indices $\nu_j$ are even, has an expansion in even powers of $\tilde{t}$.

Let us now consider an arbitrary non-zero lattice integral $I_{\nu_1 \dots \nu_N}$.
For a non-zero lattice integral $I_{\nu_1 \dots \nu_N}$ the sum of the indices $|\nu|=\nu_1+ \dots + \nu_N$ is necessarily an even integer
(if $|\nu|$ is odd the lattice integral vanishes due to the global ${\mathbb Z}_2$-symmetry).
Let $n_{\mathrm{odd}}$ be the number of odd indices $\nu_j$.
For a non-zero lattice integral $n_{\mathrm{odd}}$ is an even number and $n_{\mathrm{odd}}/2$ is an integer.
Hence, the set of lattice points with an odd index $\nu_j$ can be grouped in pairs.
Let $T$ be a forest with a minimal number of edges, such that the lattice points with odd indices $\nu_j$ are pairwise connected by this forest.
In general, $T$ will not be unique.
If $T$ has $k$ edges, the lattice integral starts at order $\tilde{t}^k$.
By the same argument as above it follows that $I_{\nu_1 \dots \nu_N}$ has an expansion in $\tilde{t}^{k}$, $\tilde{t}^{k+2}$, $\tilde{t}^{k+4}$, etc..


\section{Numerical results}
\label{sect:numerical_results}

In this section we give results for $D \in \{1,2,3,4\}$ space-time dimensions.
We always take $L=2$ lattice points in any dimension.
Our lattice consists therefore of $N=2^D$ lattice points in $D$ space-time dimensions.
We label the lattice points such that the lattice point $x_1$ has coordinates $(0,0,\dots,0)$ and
the lattice point $x_2$ has coordinates $(1,0,\dots,0)$.
The points $x_1$ and $x_2$ are time-like separated.
For all examples we set the mass equal to $m=1$.
Our main example is the two-point correlation function
\bq
 G_{110\dots0} & = & \frac{I_{110\dots0}}{I_{000\dots0}}.
\eq
We emphasize that by solving the system of differential equations we simultaneously obtain all lattice integrals
from the spanning set $\{I_1, \dots, I_{\NO}\}$.

There are two technical parameters: The number of digits $N_{\mathrm{digits}}$ used internally and
the order $N_{\mathrm{truncation}}$ at which we truncate 
the series on the right-hand side of eq.~(\ref{series_expansion_ttilde}).
We choose these parameters large enough such that they don't influence the desired precision of the final result.
Typical values are $N_{\mathrm{digits}}=100$ and $N_{\mathrm{truncation}}=200$ for a desired precision of $N_{\mathrm{desired}}=6$
digits in the case of a Minkowskian lattice in four space-time dimensions for $\lambda \approx 1$.
As a rule of thumb, lower values of $N_{\mathrm{digits}}$ and $N_{\mathrm{truncation}}$ can be used
in the Euclidean case and/or in lower dimensions.
On the other hand, lower values of the coupling $\lambda$ might require higher values 
of $N_{\mathrm{digits}}$ and $N_{\mathrm{truncation}}$.
To give an example, the value $\lambda = 0.3$ requires $N_{\mathrm{truncation}}=530$ terms 
in the $\tilde{t}$-expansion to reach an accuracy of $N_{\mathrm{desired}}=6$ digits.
There are cancellations of about twenty digits between the individual terms in the $\tilde{t}$-expansion.
In addition, there are cancellations in computing the next term in the $\tilde{t}$-expansion from the previous ones.

\begin{figure}
\begin{center}
\includegraphics[scale=0.6]{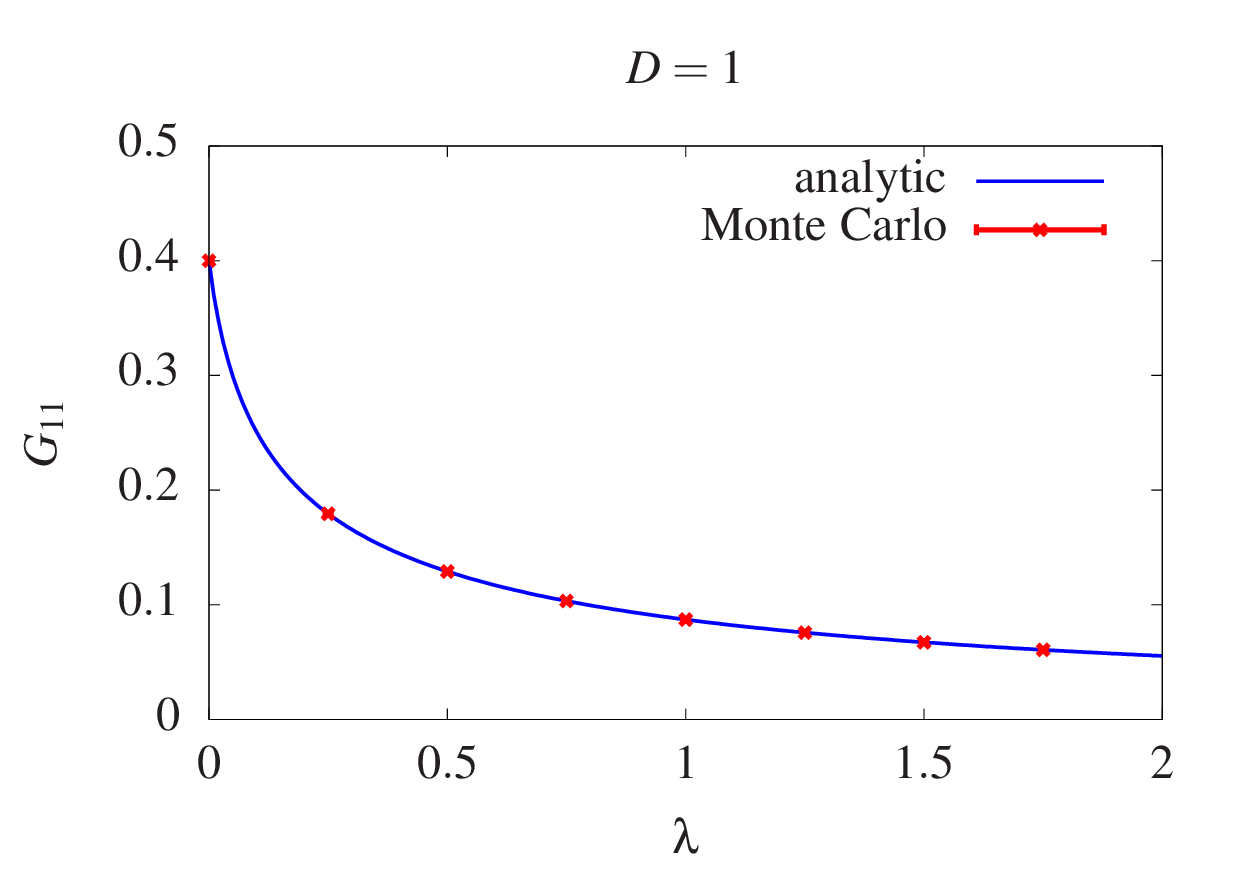}
\includegraphics[scale=0.6]{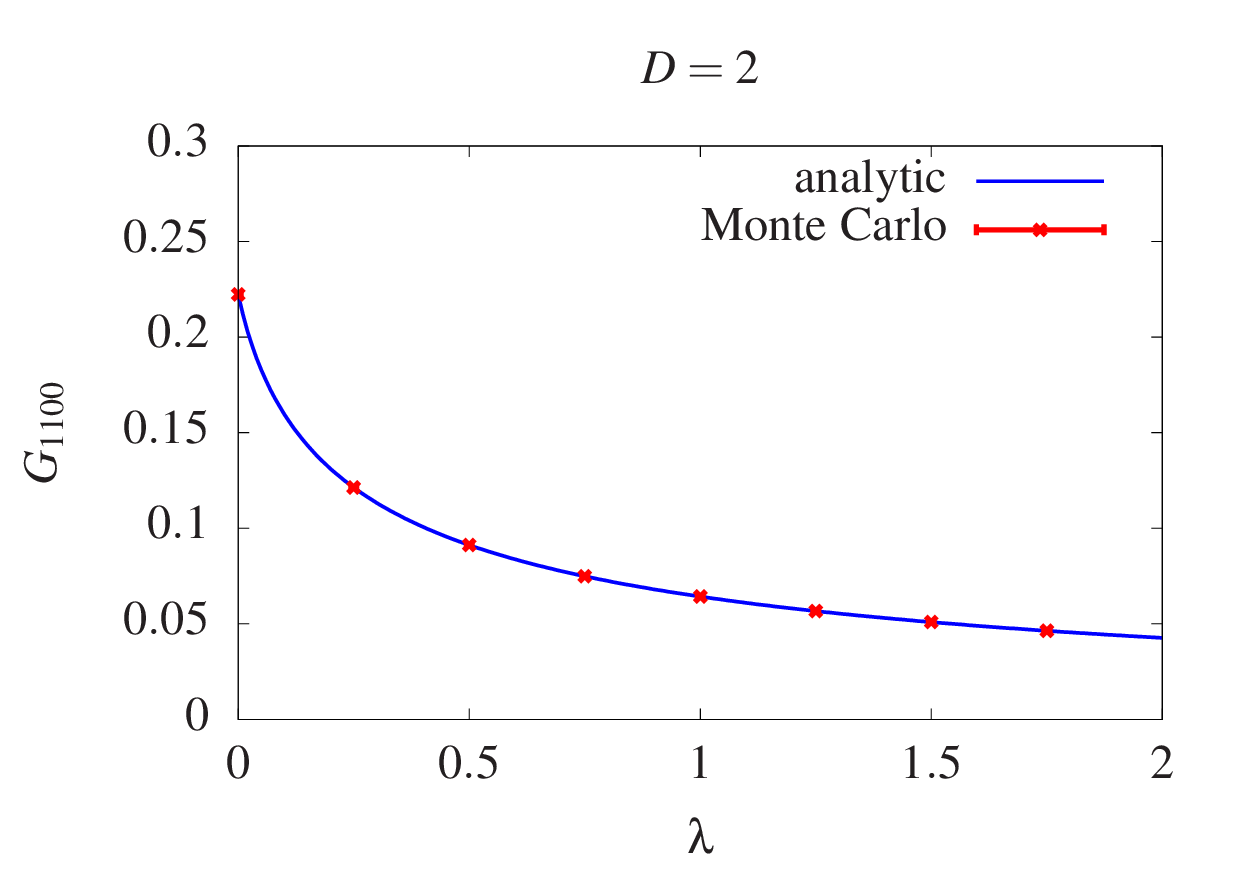}
\includegraphics[scale=0.6]{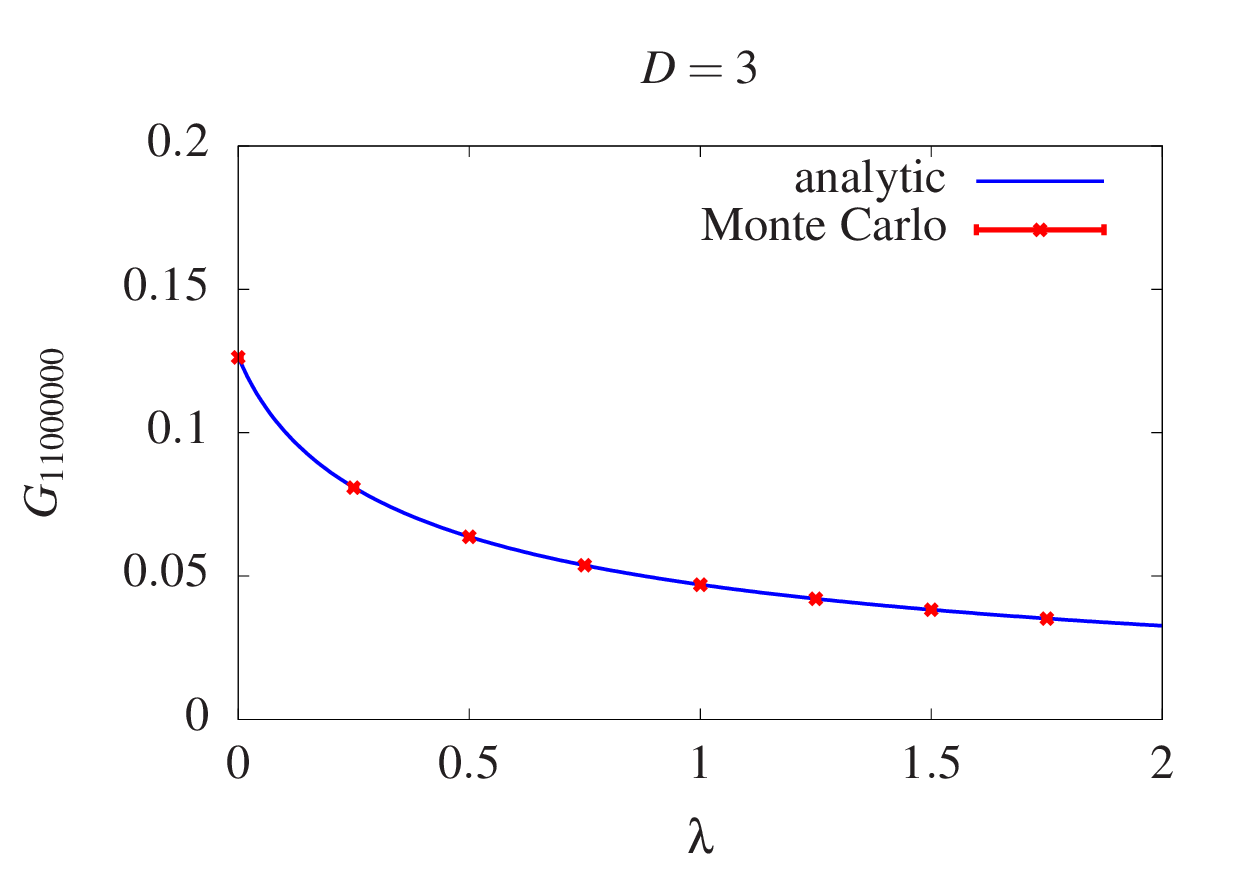}
\includegraphics[scale=0.6]{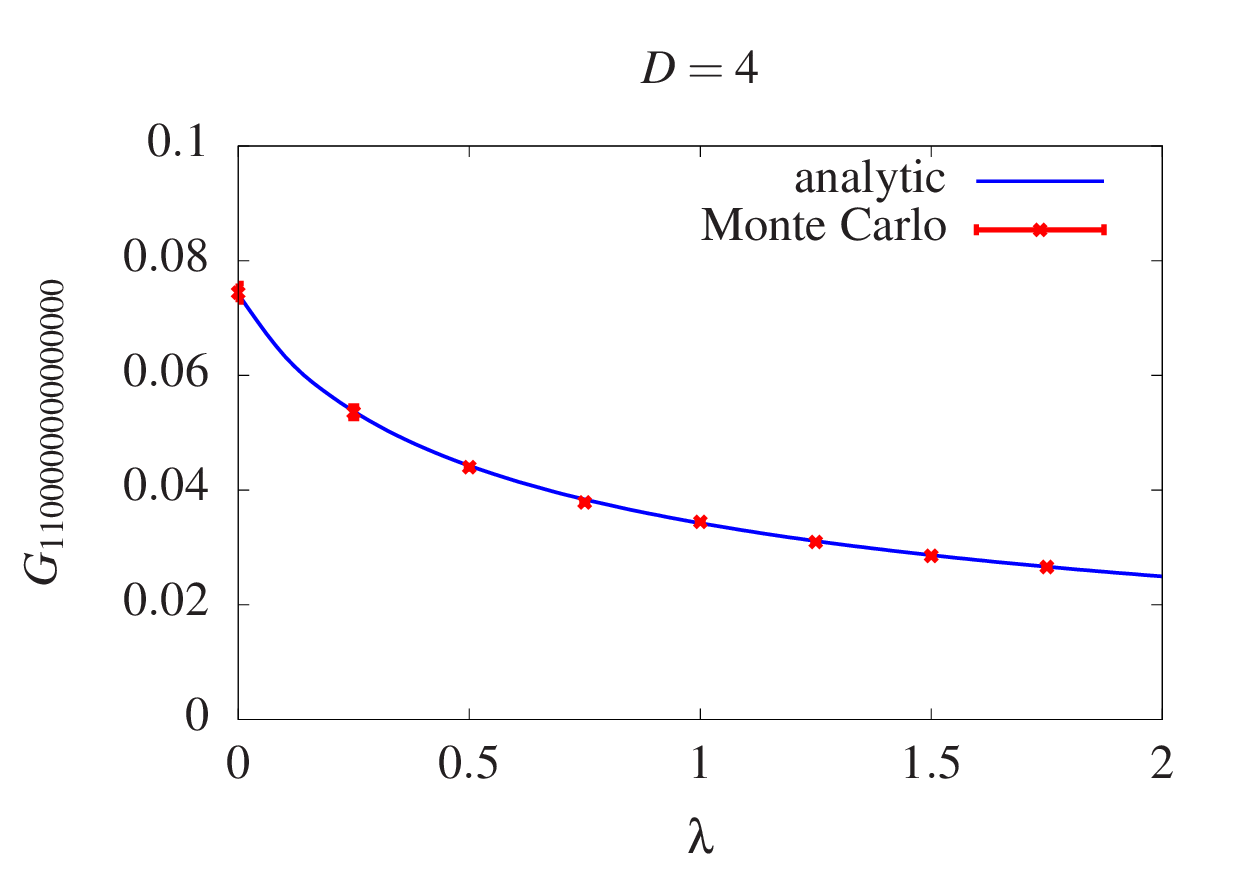}
\end{center}
\caption{
The Euclidean correlation function $G_{110\dots0}$ as a function of the coupling $\lambda$ 
for one, two, three and four space-time dimensions (upper left to lower right).
}
\label{fig_G11_euclidean}
\end{figure}

We first validate our approach for the Euclidean case.
In fig.~\ref{fig_G11_euclidean} we show the Euclidean correlation function $G_{110\dots0}$
as a function of the coupling $\lambda$ for $D\in\{1,2,3,4\}$ space-time dimensions.
For comparison we also show results obtained from Monte Carlo integration.
For the Monte Carlo integration we used the {\tt VEGAS} algorithm \cite{Lepage:1978sw,Lepage:1980dq}.
We observe perfect agreement.

\begin{figure}
\begin{center}
\includegraphics[scale=0.6]{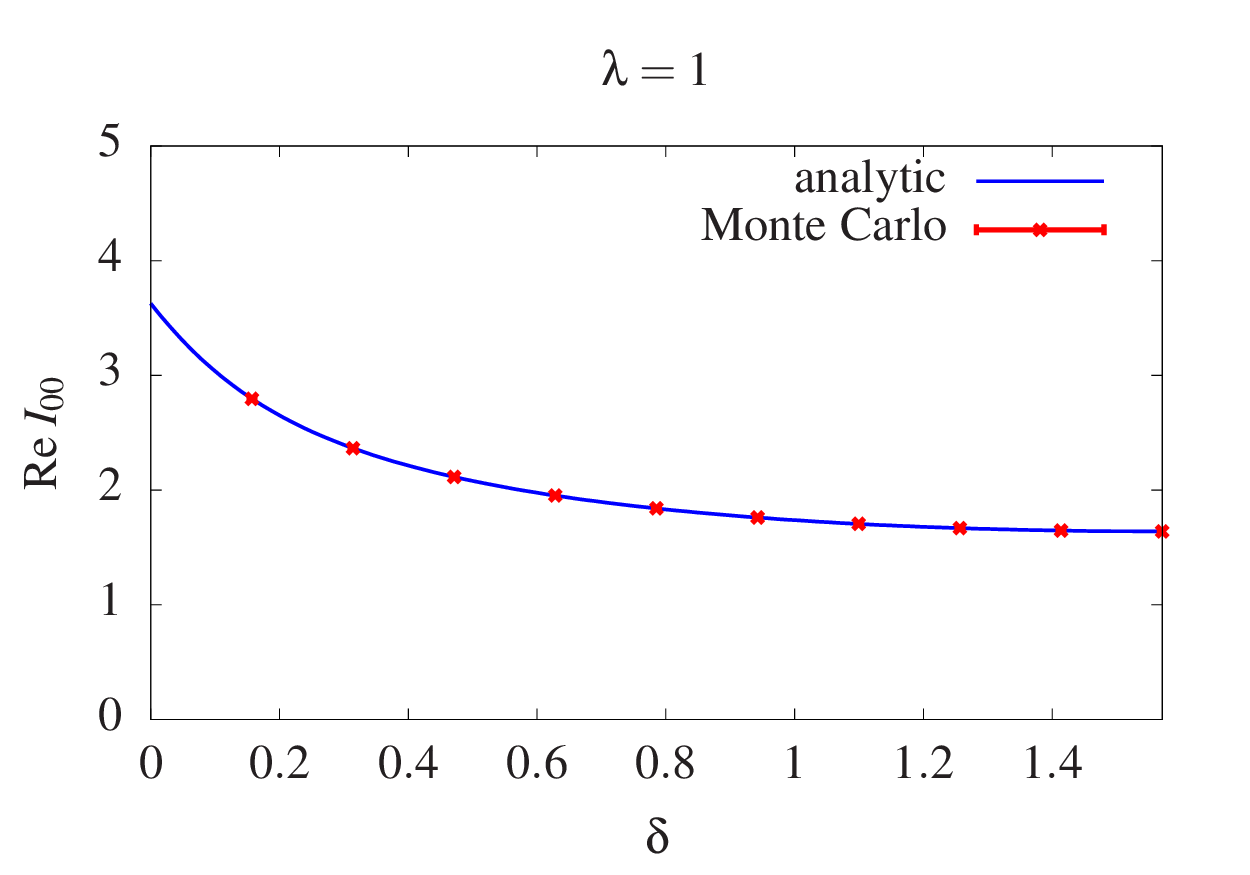}
\includegraphics[scale=0.6]{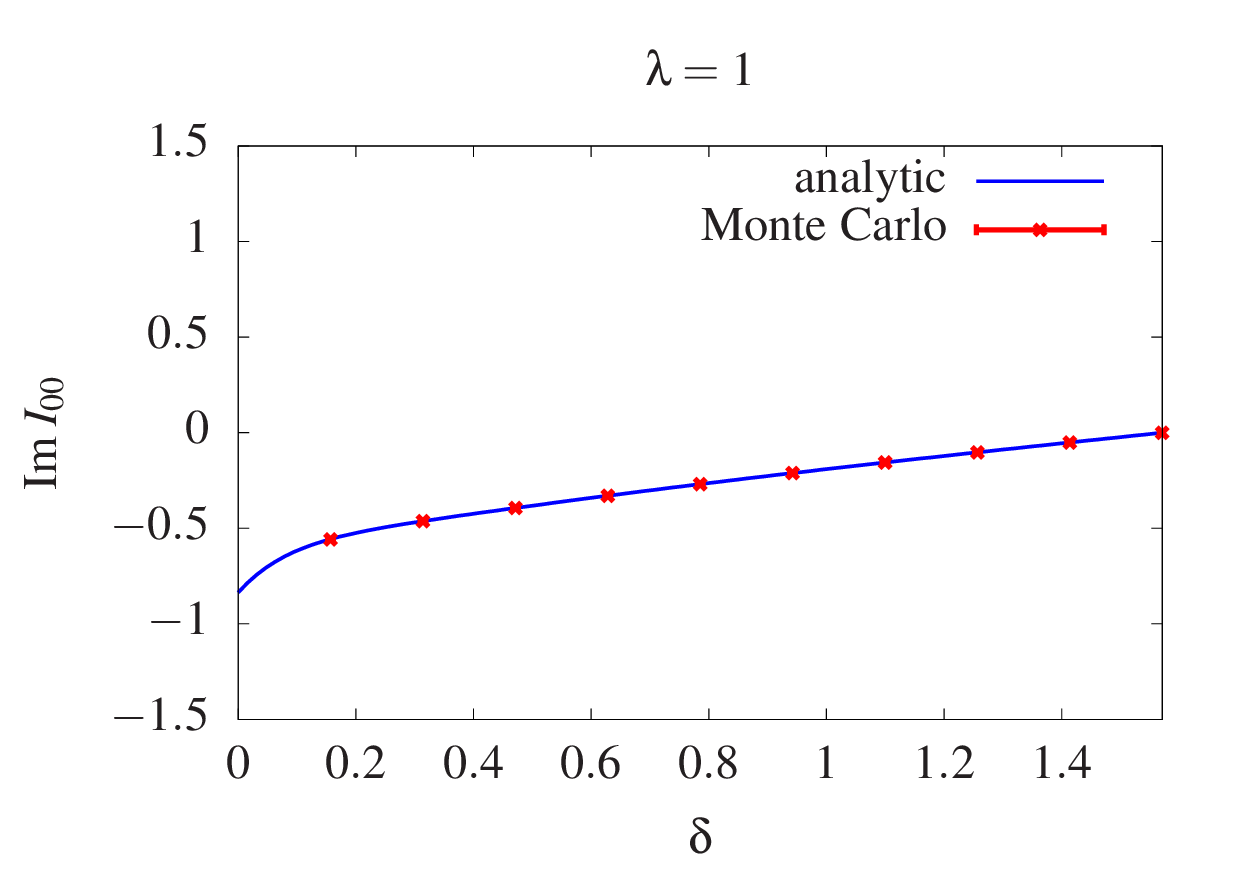}
\end{center}
\caption{
The lattice integral $I_{00}$ in $D=1$ space-time dimensions as a function of the Wick rotation angle $\delta \in [0,\frac{\pi}{2}]$ for $\lambda=1$.
The left plot shows the real part, the right plot shows the imaginary part.
}
\label{fig_I00_delta}
\end{figure}

We then validate our approach for an arbitrary Wick rotation angle $\delta$.
In fig.~\ref{fig_I00_delta} we show the real and imaginary part of the lattice integral $I_{00}$
in $D=1$ space-time dimensions as a function of the Wick rotation angle $\delta \in [0,\frac{\pi}{2}]$ for $\lambda=1$.
For comparison we also show results obtained from Monte Carlo integration.
Again, we observe perfect agreement.

\begin{figure}
\begin{center}
\includegraphics[scale=0.6]{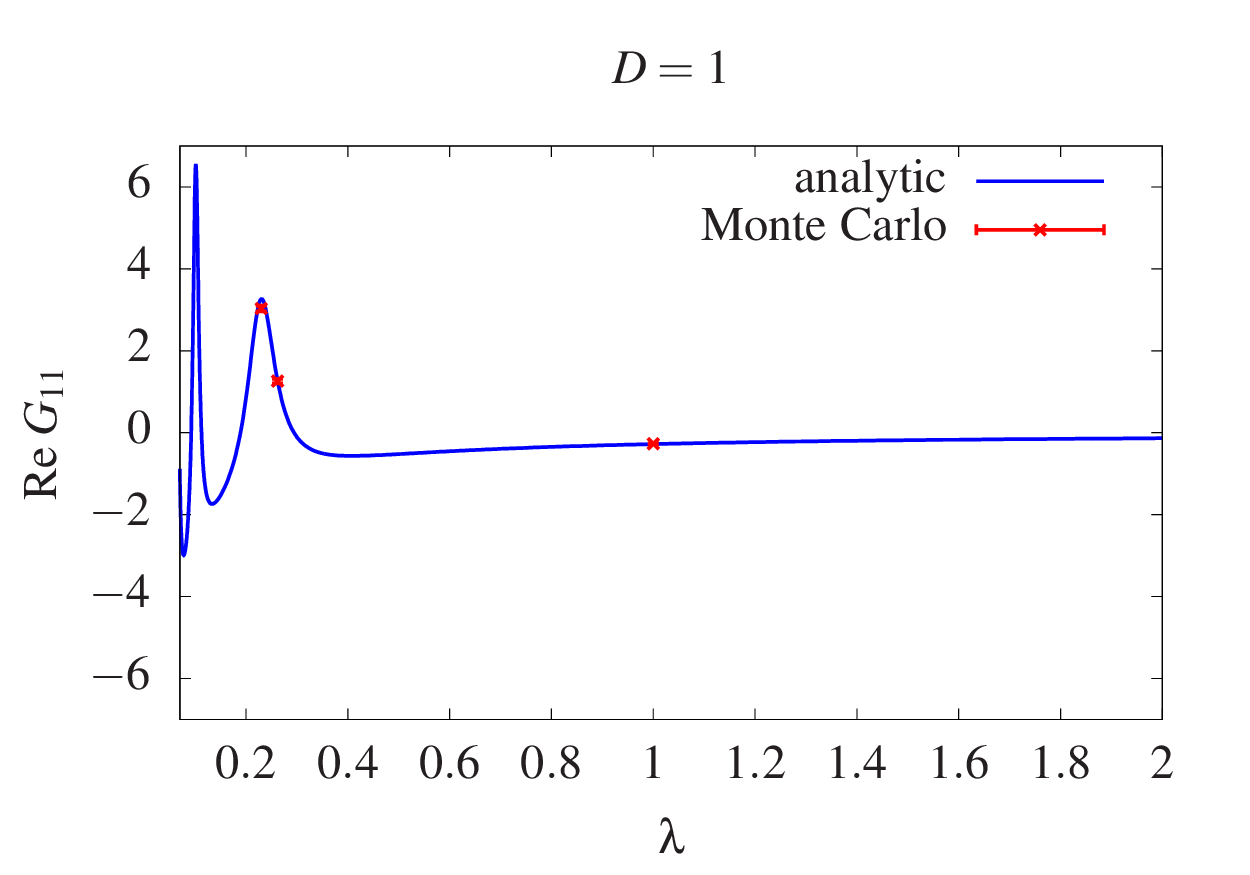}
\includegraphics[scale=0.6]{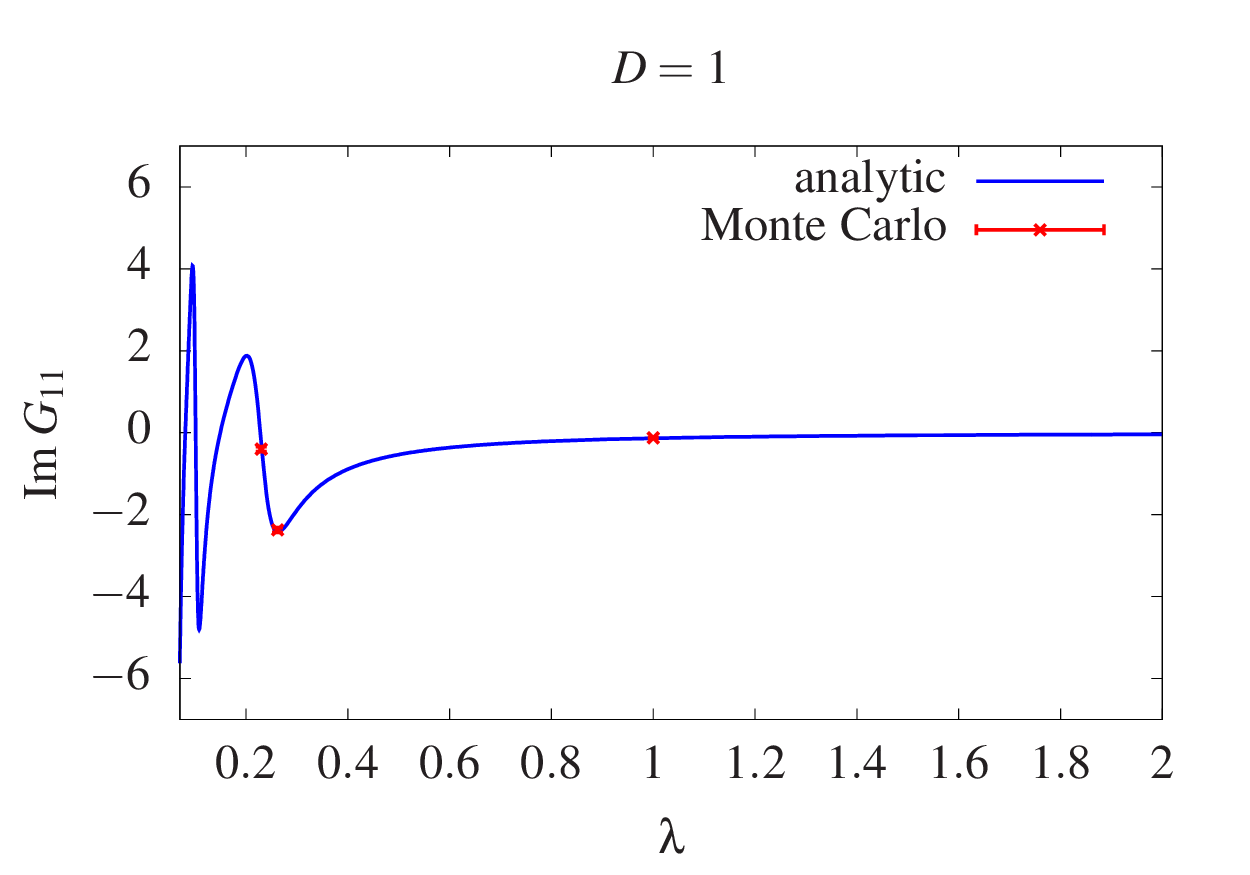}
\end{center}
\caption{
The Minkowskian correlation function $G_{11}$ in one space-time dimension as a function of the coupling $\lambda \in [0.07,2]$.
The left plot shows the real part, the right plot shows the imaginary part.
}
\label{fig_Minkowskian_D_eq_1}
\end{figure}

\begin{figure}
\begin{center}
\includegraphics[scale=0.6]{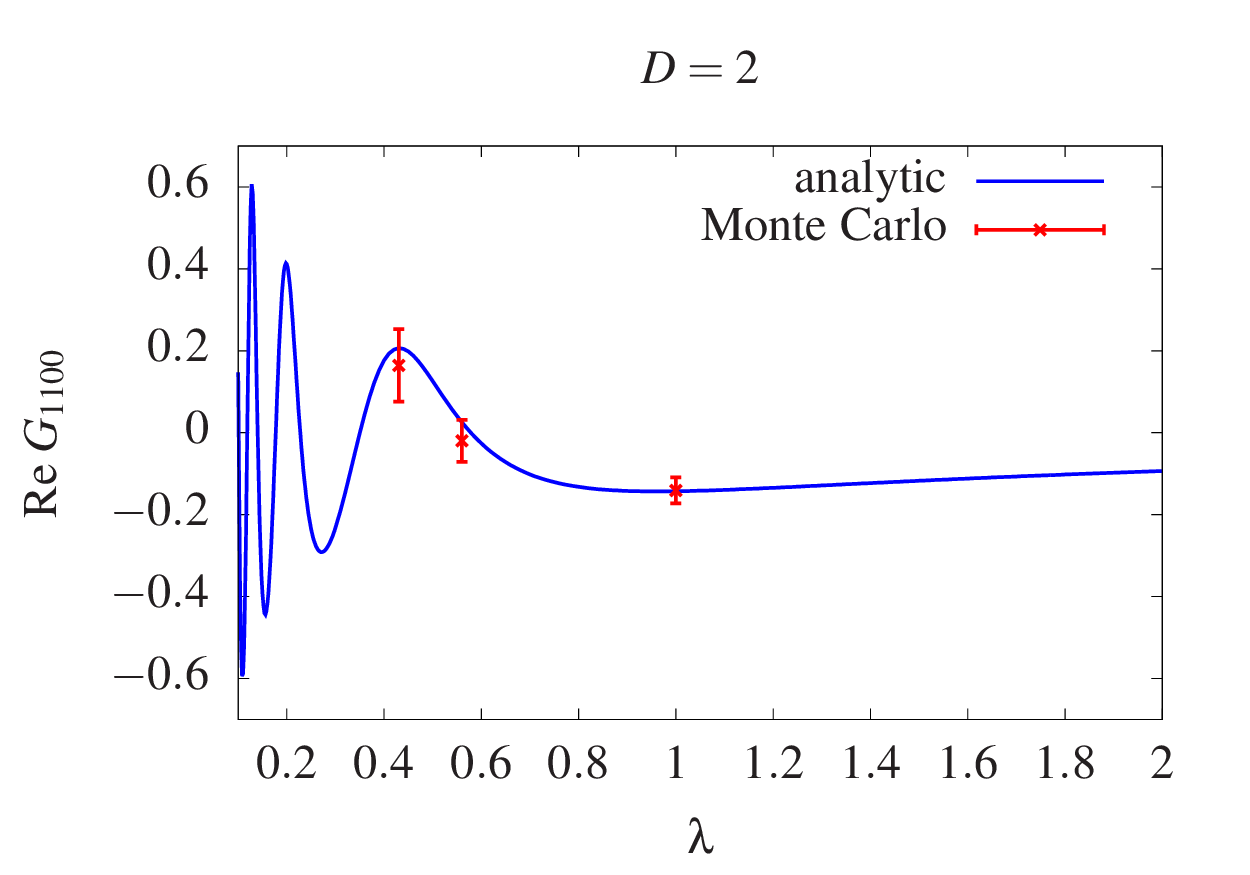}
\includegraphics[scale=0.6]{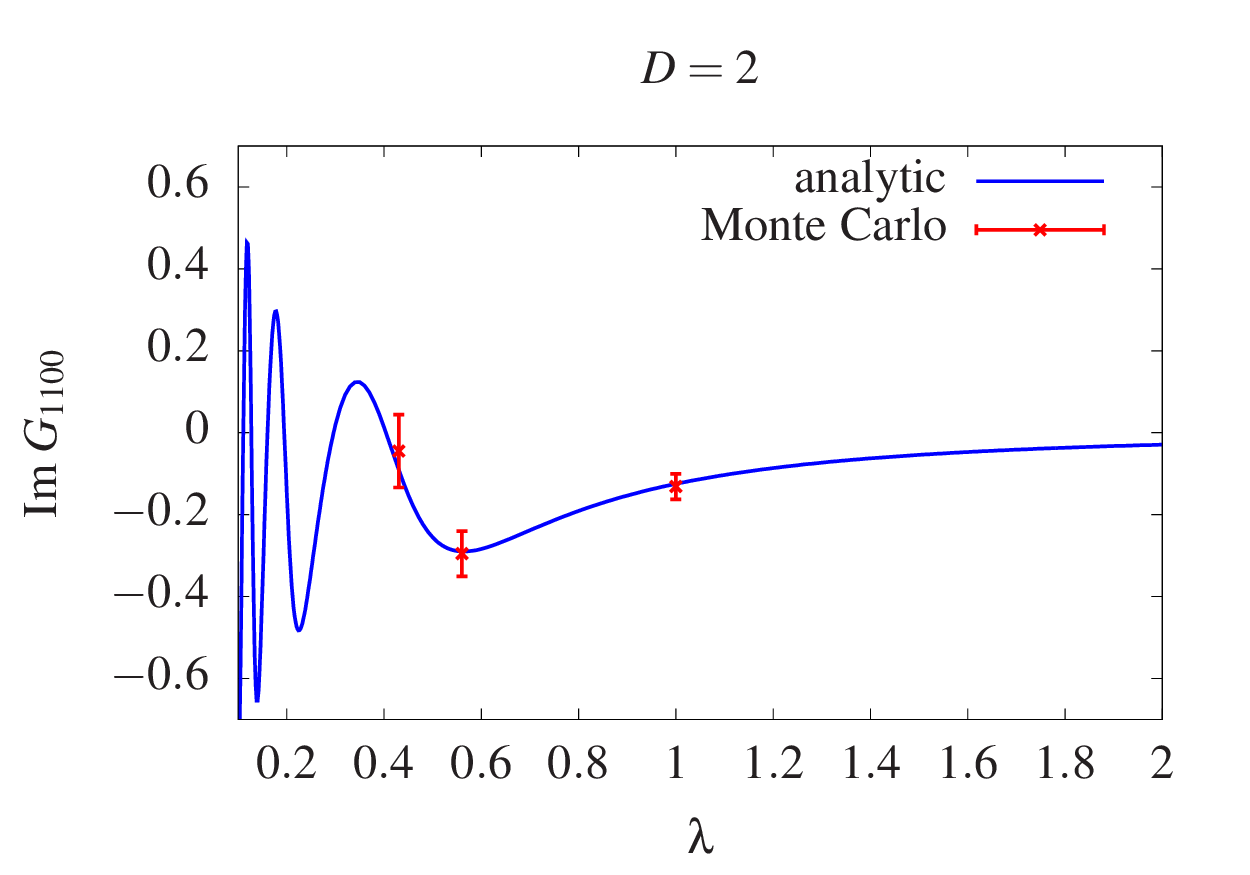}
\end{center}
\caption{
The Minkowskian correlation function $G_{1100}$ in two space-time dimension as a function of the coupling $\lambda \in [0.1,2]$.
The left plot shows the real part, the right plot shows the imaginary part.
}
\label{fig_Minkowskian_D_eq_2}
\end{figure}

\begin{figure}
\begin{center}
\includegraphics[scale=0.6]{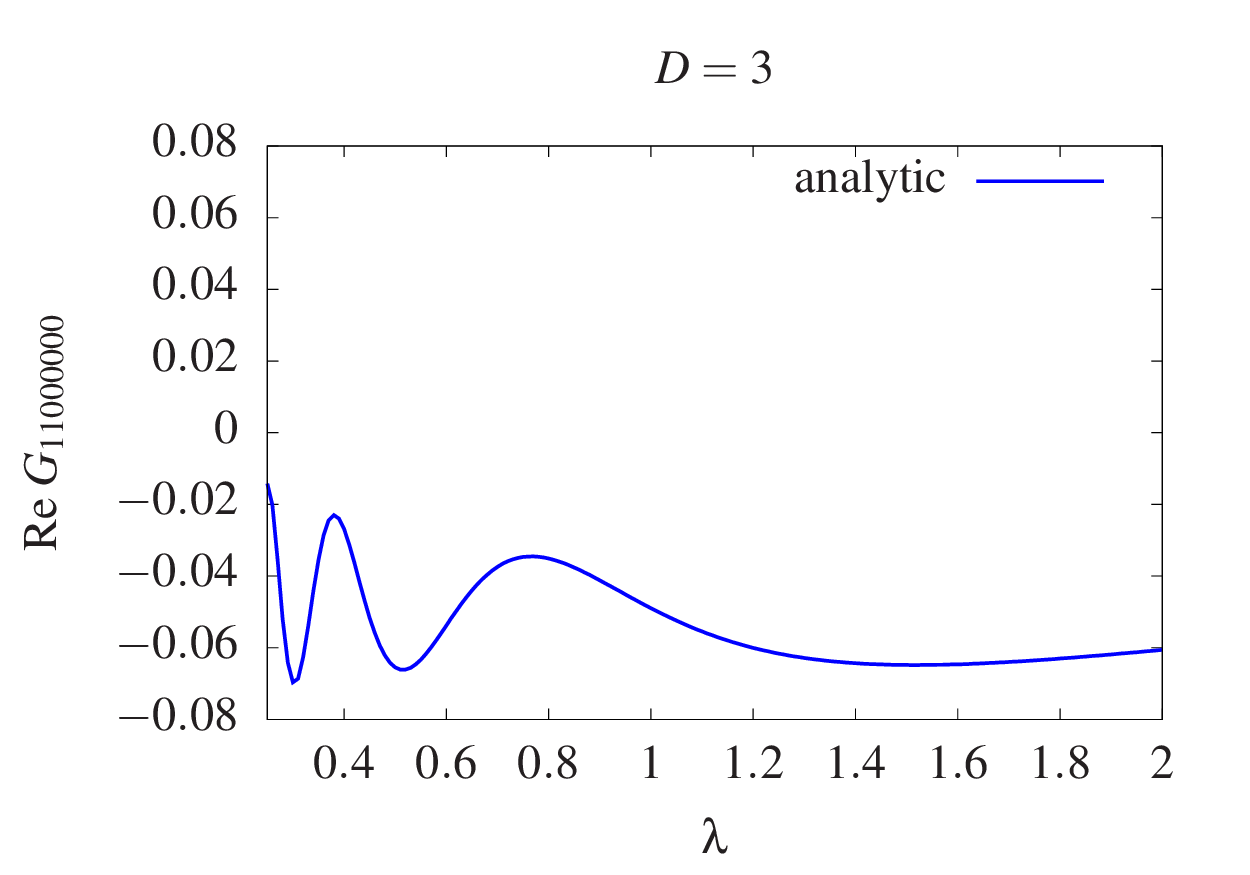}
\includegraphics[scale=0.6]{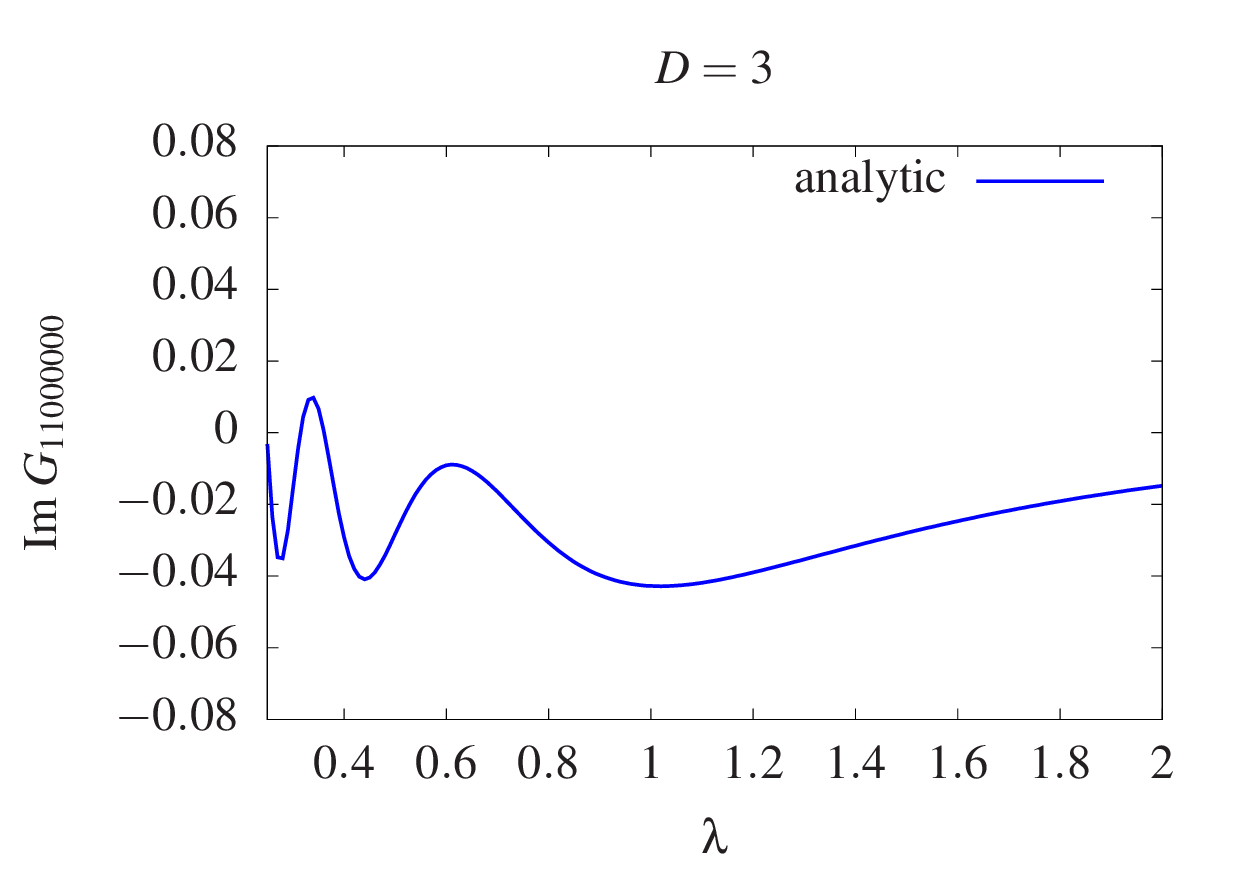}
\end{center}
\caption{
The Minkowskian correlation function $G_{11000000}$ in three space-time dimension as a function of the coupling $\lambda \in [0.25,2]$.
The left plot shows the real part, the right plot shows the imaginary part.
}
\label{fig_Minkowskian_D_eq_3}
\end{figure}

\begin{figure}
\begin{center}
\includegraphics[scale=0.6]{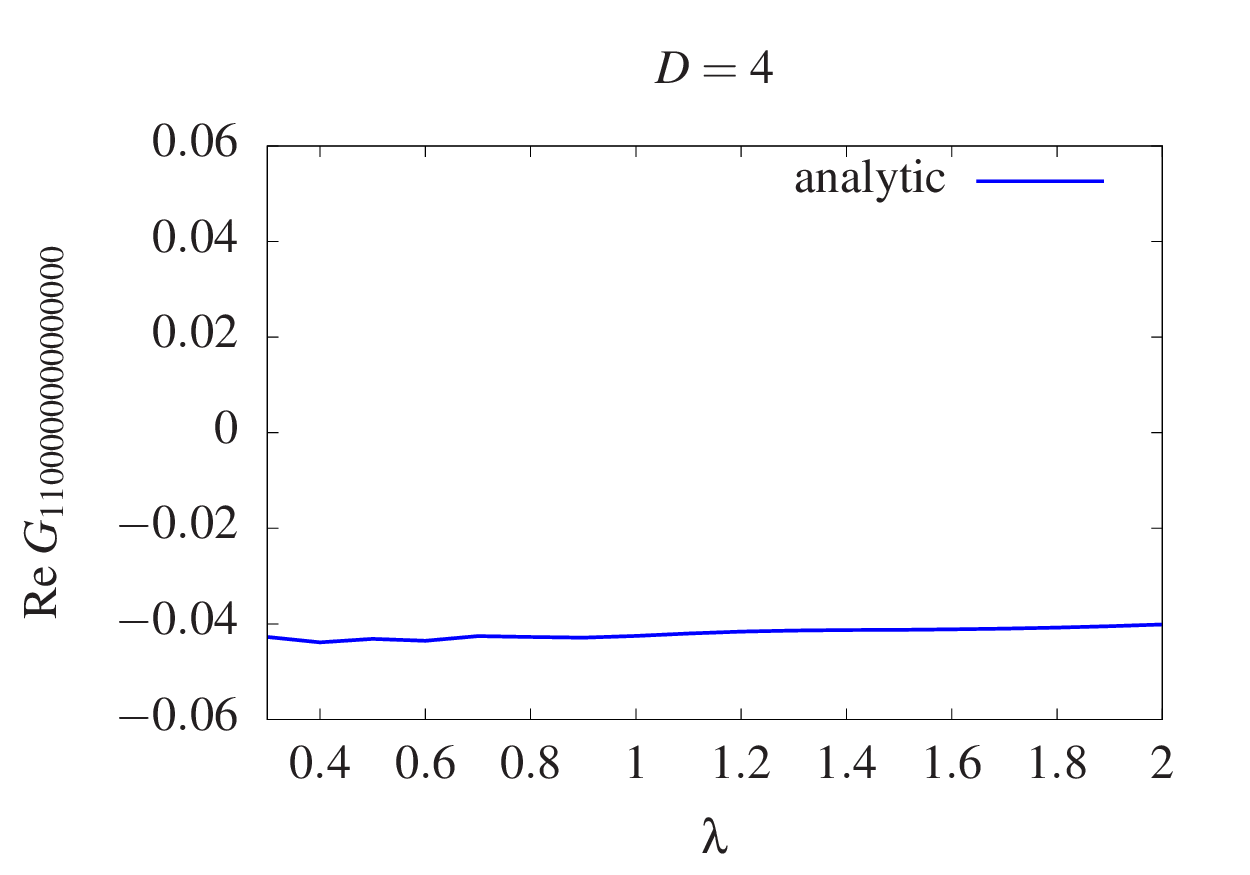}
\includegraphics[scale=0.6]{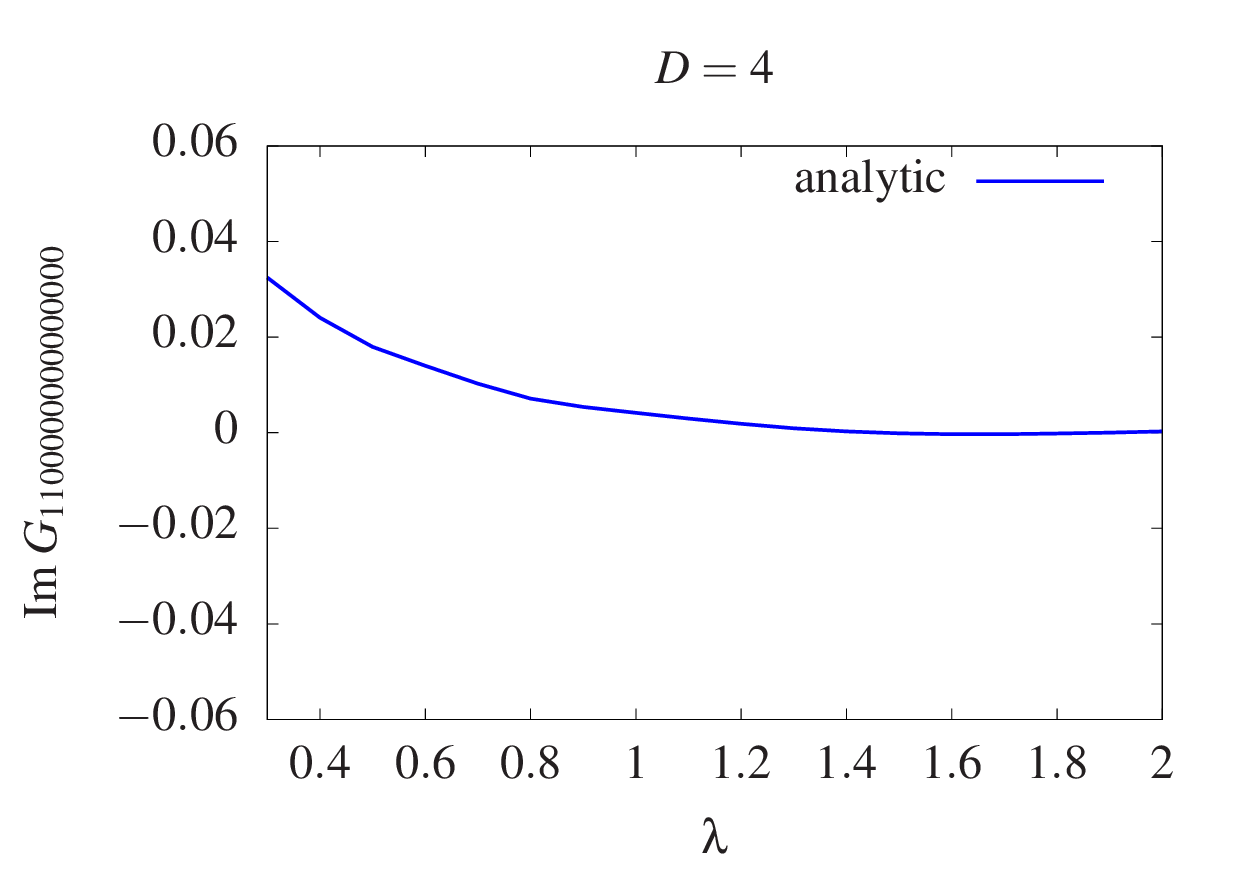}
\end{center}
\caption{
The Minkowskian correlation function $G_{1100000000000000}$ in four space-time dimension as a function of the coupling $\lambda \in [0.3,2]$.
The left plot shows the real part, the right plot shows the imaginary part.
}
\label{fig_Minkowskian_D_eq_4}
\end{figure}

\begin{figure}
\begin{center}
\includegraphics[scale=0.6]{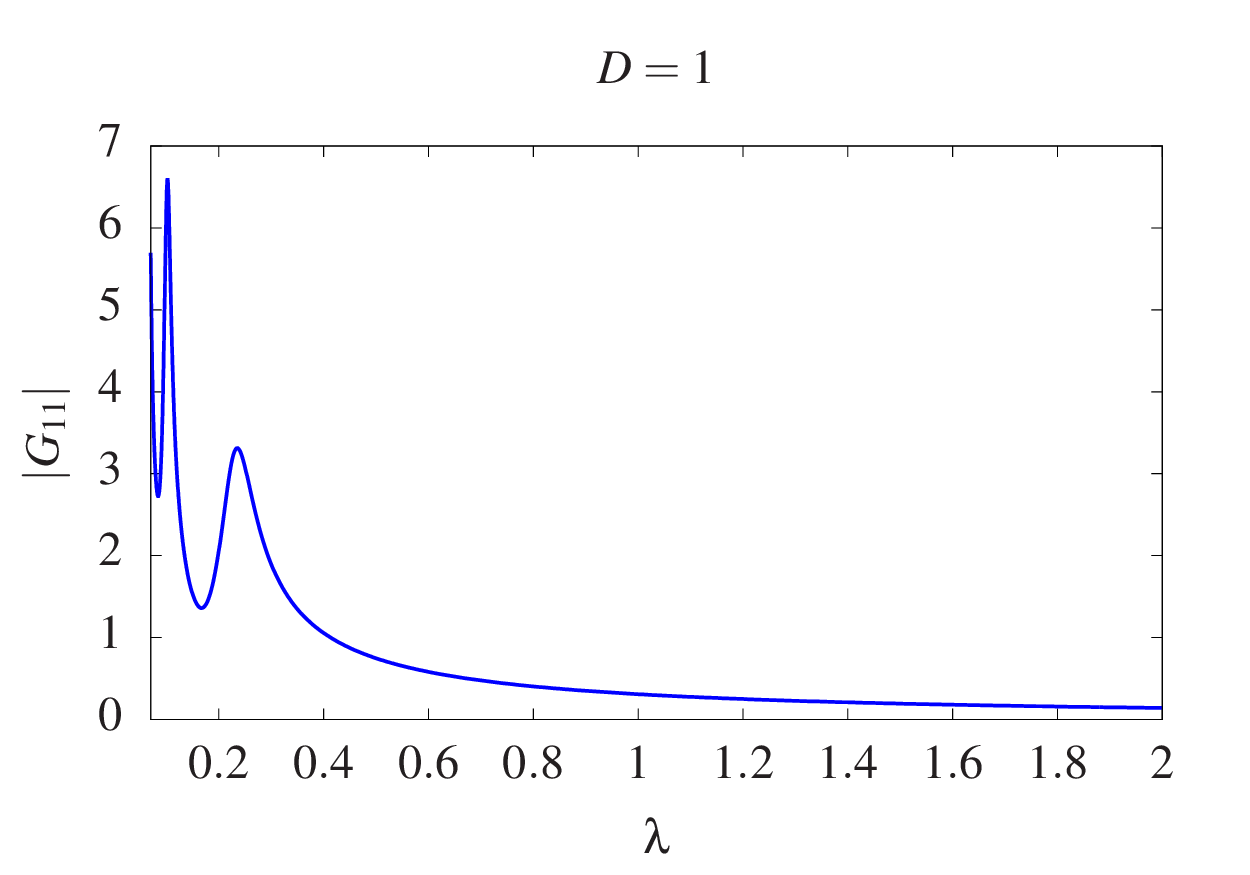}
\end{center}
\caption{
The absolute value of the Minkowskian correlation function $G_{11}$ in one space-time dimension as a function of the coupling $\lambda \in [0.07,2]$.
}
\label{fig_Minkowskian_D_eq_1_abs}
\end{figure}

Let us now turn to the Minkowskian case:
In figs.~\ref{fig_Minkowskian_D_eq_1}-\ref{fig_Minkowskian_D_eq_4}
we show the real and the imaginary part of the Minkowskian correlation function $G_{110\dots0}$
as a function of the coupling $\lambda$ for $D\in\{1,2,3,4\}$ space-time dimensions.
Contrary to the Euclidean case we observe an oscillatory behaviour as a function of the coupling $\lambda$.
The period of the oscillations becomes smaller as we go to smaller values of $\lambda$.
For this reason we don't plot them down to $\lambda=0$, but in an interval $[\lambda_0,2]$ with $0 < \lambda_0 < 2$.
This avoids the wildly oscillating part.
The oscillatory behaviour is not just a phase. 
This can be deduced from fig.~\ref{fig_Minkowskian_D_eq_1_abs},
where we plot the absolute value of the 
Minkowskian correlation function $G_{11}$ in one space-time dimension as a function of the coupling.
We see that the absolute value is also oscillating.
In one and two space-time dimensions we also include in the plots of figs.~\ref{fig_Minkowskian_D_eq_1}-\ref{fig_Minkowskian_D_eq_2}
results from Monte Carlo integration. These are obtained as follows: We use a small, but finite Wick rotation angle $\delta$.
As we decrease $\delta$, the integrand will oscillate more and the Monte Carlo integration error will increase (if we keep the number of integrand
evaluations constant). 
We study the dependence on the Wick rotation angle until the systematic error from a finite non-zero Wick rotation angle is smaller than
the statistical error from the Monte Carlo integration.
As we increase the number of space-time dimensions this becomes more expensive.
(This is expected and one of the motivations for the present research.)
The results in one and two space dimensions from Monte Carlo integration agree with the analytical results.

\subsection{Comparison with perturbation theory}

In this section we compare the exact results with results obtained from perturbation theory.
We recall that with our convention the interaction term is given by (see eq.~(\ref{interaction_term}))
\bq
 S^{(4)}
 & = &
 \frac{i \lambda}{\alpha}
 \sum\limits_{x \in \Lambda}
 \phi_x^4,
\eq
i.e. without a factor $1/24$. This convention is convenient for our purposes, as it avoids powers of $24$ in the
differential equation.
Within perturbation theory one often uses the convention
\bq
 S^{(4)}
 & = &
 \frac{i \hat{\lambda}}{24 \alpha}
 \sum\limits_{x \in \Lambda}
 \phi_x^4.
\eq
This is convenient in perturbation theory, as the factor $1/24$ compensates the $4!$ possibilities to connect
four propagators to the vertex.
The relation between the two conventions is $\hat{\lambda} = 24 \lambda$, and $\lambda=0.04$ corresponds to 
$\hat{\lambda} = 0.96$.

\begin{figure}
\begin{center}
\includegraphics[scale=0.6]{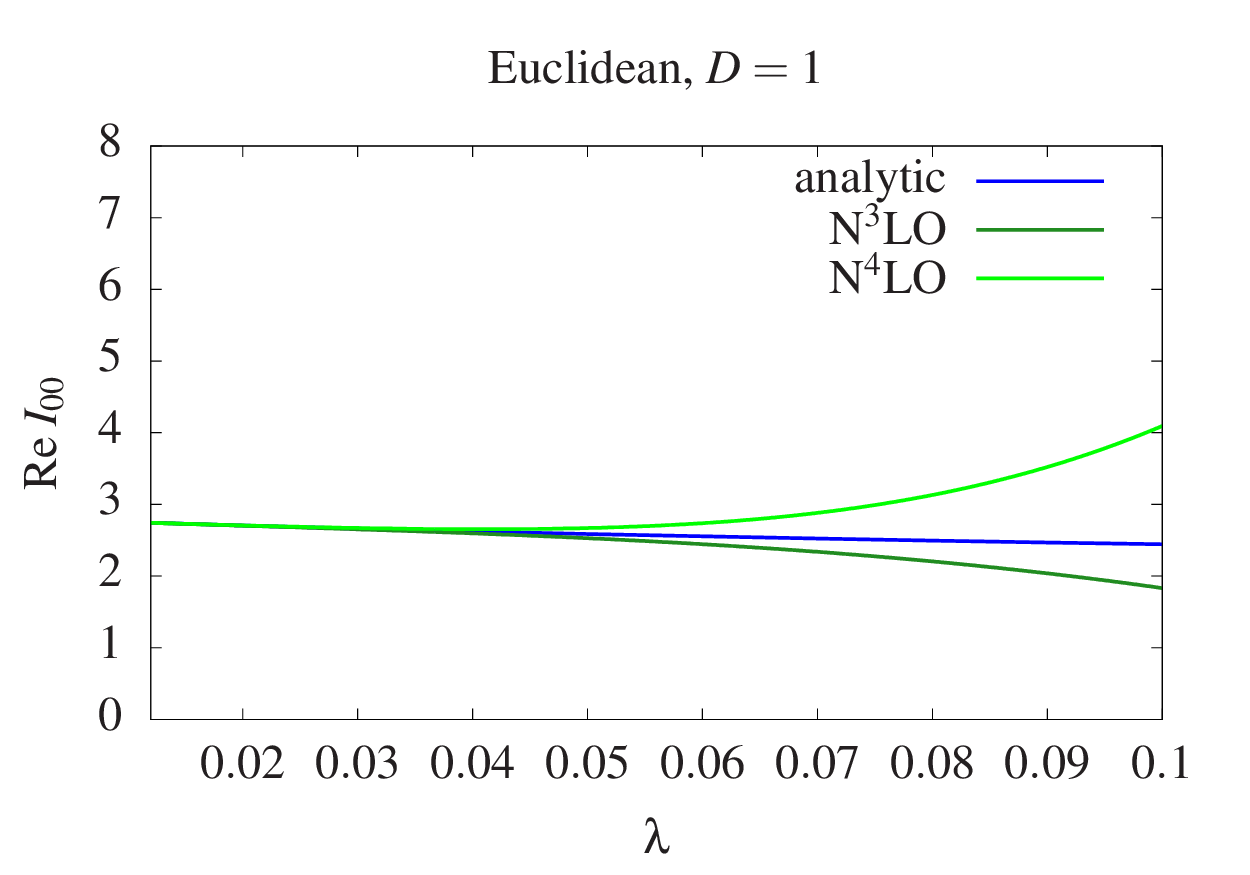}
\includegraphics[scale=0.6]{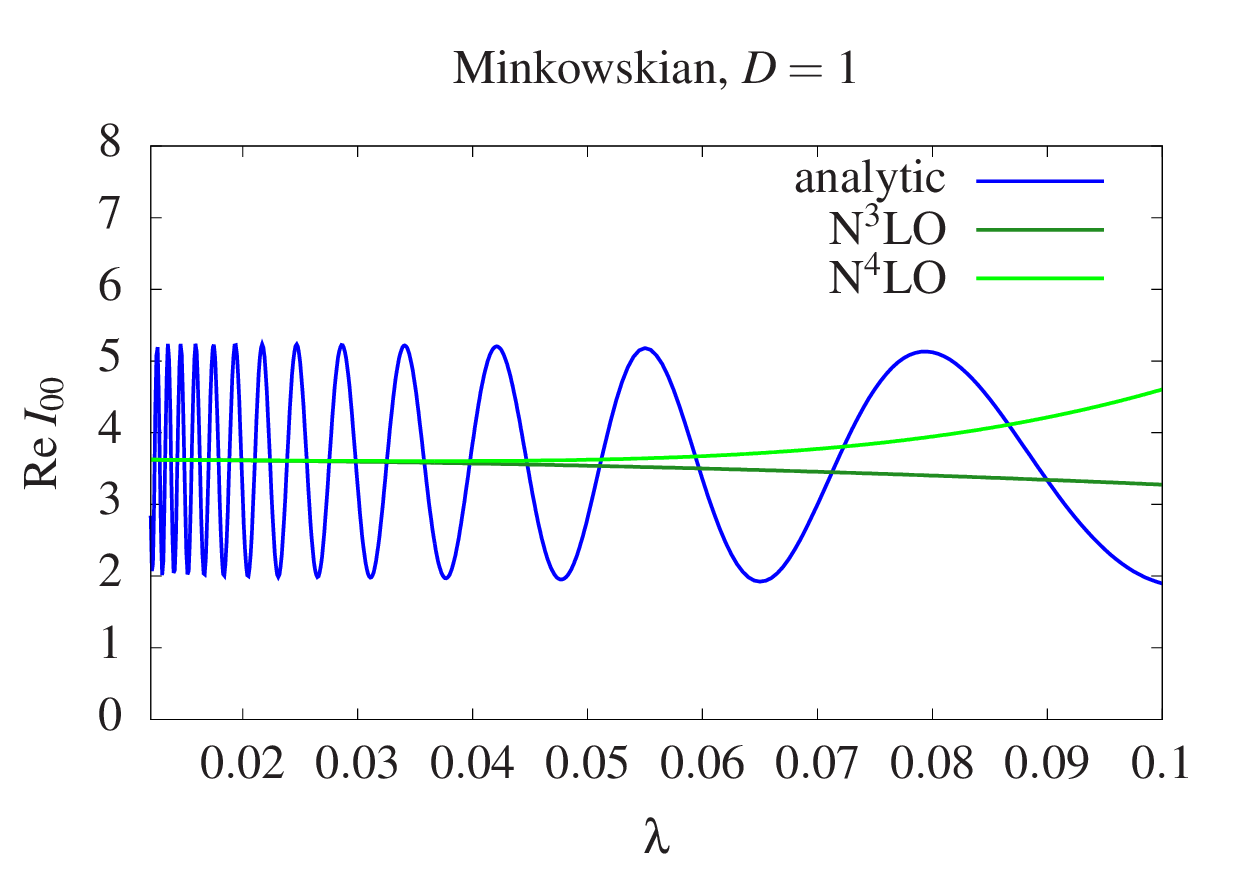}
\end{center}
\caption{
The left plot shows the Euclidean lattice integral $I_{00}$ as a function of the coupling $\lambda$ in the range $[0.012,0.1]$.
The green line shows the analytic result, the cyan line the perturbative N${}^3$LO result, 
the blue line shows the perturbative N${}^4$LO result.
For $\lambda \lesssim 0.04$ the results agree.
The right plot shows the real part of Minkowskian lattice integral $I_{00}$ as a function of the coupling $\lambda$ in the range $[0.012,0.1]$.
The green line shows the analytic result, the cyan line the perturbative N${}^3$LO result, 
the blue line shows the perturbative N${}^4$LO result.
The perturbative results do not reproduce the oscillations.
}
\label{fig_I_00_lambda_Minkowskian_D_eq_1_real}
\end{figure}

We now study the lattice integrals for small values of $\lambda$.
We compare the analytic result with perturbation theory.
The left plot of fig.~\ref{fig_I_00_lambda_Minkowskian_D_eq_1_real} shows the Euclidean lattice integral $I_{00}$ 
in one space-time dimension as a function of the coupling $\lambda \in [0.012,0.1]$.
We show the exact result together with the perturbative N${}^3$LO and N${}^4$LO approximations.
For $\lambda \lesssim 0.04$ the three results agree reasonably.
We recall that the perturbative series is just an asymptotic series and it is expected that the perturbative results
will diverge from the exact results for larger values of the coupling $\lambda$.

Let us now consider the Minkowskian case.
The right plot of fig.~\ref{fig_I_00_lambda_Minkowskian_D_eq_1_real} 
shows the real part of the Minkowskian lattice integral $I_{00}$ 
in one space-time dimension as a function of the coupling $\lambda \in [0.012,0.1]$.
We show again the exact result together with the perturbative N${}^3$LO and N${}^4$LO approximations.
The exact result is oscillating, the period of the oscillations is decreasing with decreasing $\lambda$.
This behaviour is not reproduced by the perturbative results, which are slowly varying functions of the coupling $\lambda$.

This can be understood as follows: Let us go back to the Euclidean case.
A simplified model for the result in the Euclidean case is
\bq
\label{simplified_model}
 A + B e^{-\frac{c}{\lambda}},
 \;\;\;\;\;\; c > 0,
\eq
where $A$ (and $B$) are slowly varying functions of $\lambda$. 
For a qualitative discussion we may treat them as constants.
We stress that the simplified function in eq.~(\ref{simplified_model}) 
serves only to discuss the qualitative features of the solution.
At small coupling, the second term is exponentially suppressed and perturbation theory accurately predicts $A$.
For small coupling $\lambda$ the magnitude of $B$ is irrelevant, as it is multiplied by an exponentially small factor. 

From section~\ref{sect:efficiency_improvements} 
we see that the dependence on the coupling (apart from a prefactor) is essentially given by the combination
\bq
 \chi^2 & = & \frac{i \mu^2}{8 \alpha \lambda}.
\eq
In the Minkowskian case our simplified model becomes
\bq
 A + B e^{i \frac{c}{\lambda}},
 \;\;\;\;\;\; c > 0.
\eq
The difference is the additional factor $i$ in the exponent.
The second term is now oscillatory.
Perturbation theory again accurately predicts $A$, but gives no information on the second term.
The accuracy of a perturbative calculation is therefore limited by the amplitude $B$ of the oscillations.
This is expected \cite{Chibisov:1996wf,Blok:1997hs}.

Let us now return from the simplified model to the exact result.
With the exact result at hand we may quantify the accuracy of perturbative approximations.
To give an example, the exact value at $\lambda = 0.0286$ is
\bq
 \mathrm{Re} \; I_{00} & = & 5.22334
\eq
Perturbation theory gives
\bq
 \mathrm{Re} \; I_{00}^{\mathrm{N}^3\mathrm{LO}}  & = & 3.59881,
 \nonumber \\
 \mathrm{Re} \; I_{00}^{\mathrm{N}^4\mathrm{LO}}  & = & 3.60770.
\eq
Adding additional higher orders within perturbation theory will not change the picture significantly.
The error of the perturbative result is
\bq
 \left| \mathrm{Re} \; I_{00} - \mathrm{Re} \; I_{00}^{\mathrm{N}^3\mathrm{LO}} \right|
 & = &
 1.62453.
\eq
The precision of the perturbative result is usually estimated from the last calculated term.
This gives
\bq
 \left| \mathrm{Re} \; I_{00}^{\mathrm{N}^4\mathrm{LO}} - \mathrm{Re} \; I_{00}^{\mathrm{N}^3\mathrm{LO}} \right|
 & = &
 0.00889.
\eq
We see that in the Minkowskian case there are sizeable contributions to the exact result not predicted by 
perturbation theory.

We also stress that in the Minkowskian case we cannot use perturbative results as boundary values for a differential
equation in $\lambda$.
As they are missing non-perturbative contributions, they are simply not the right boundary values.


\section{Conclusions}
\label{sect:conclusions}

In this paper we reported on the analytic calculation of lattice correlation functions
for $\phi^4$-theory on a lattice in one, two, three or four space-time dimensions with 
either Euclidean or Minkowskian signature.
The lattice correlation functions have been calculated by the method of differential equations.
Integration-by-parts identities lead to a finite system of first-order differential equations.
The mathematical framework underlying integration-by-parts identities is twisted cohomology.
We systematically investigated the interplay between twisted cohomology 
and the symmetries of the twist function.
Symmetries reduce significantly the size of the system of differential equations.

The lattice correlation functions have a power series expansion in $1/\sqrt{\lambda}$, 
where $\lambda$ is the coupling. We showed that this series is convergent for all non-zero values of $\lambda$.
We also showed that a slightly modified expansion is better suited for numerical evaluations.
At small coupling we quantified the accuracy of perturbative approximations.

As an outlook to the near future we expect that slightly larger lattice sizes for $\phi^4$-theory are feasible,
as on the one hand there are additional optimisations, which we haven't implemented 
and on the other hand we may enlarge the computing resources by going from a single PC to a computer cluster.
On a medium time scale we do not exclude the possibility that algorithmic improvements
lead to even larger lattices.

Furthermore, we are very interested in transferring the ideas and methods of this paper from scalar $\phi^4$-theory to 
Yang-Mills theory.

\subsection*{Acknowledgements}

We would like to thank Harvey Meyer and Carlo Pagani for useful discussions.
S.W. would also like to thank Andr\'e Hoang for useful discussions.


\begin{appendix}

\section{Efficiency}
\label{sect:efficiency}

In this appendix we discuss how to implement efficiently the reduction of eq.~(\ref{reduction_to_basis_integral})
\bq
 \left\langle \Phi \left| {\mathbb R}^N \right. \right\rangle
 & = &
 \sum\limits_{i=1}^{\NF} c_i \left\langle e_i \left| {\mathbb R}^N \right. \right\rangle
\eq
to elements of a spanning set of lattice integrals.
We assume that $\Phi$ is given by
\bq
 \Phi & = &
 \hat{\Phi}\left(\phi\right) d^N\phi,
\eq
where $\hat{\Phi}(\phi)$ is a polynomial in $\phi_1,\dots,\phi_N$.
In the following we will use the convention that for any $N$-form $\Phi$ we denote by $\hat{\Phi}$
the function obtained by stripping $d^N\phi$ off.
We are interested in the situation, where the dimension $\NF$ of the twisted cohomology group is of 
the order of ${\mathcal O}(10^7)$ and the number $\NO$ of elements in the spanning set of lattice integrals
is of the order ${\mathcal O}(10^4)$.

We set
\bq
 f_j & = & \frac{i \alpha}{4 \lambda} \omega_{x_j}.
\eq
The prefactor ensures that the leading term of $f_j$ is normalised to one:
\bq
\label{Groebner_element}
 f_j & = &  \phi_{x_j}^3 + l_j.
\eq
The lower terms $l_j$ are linear in the $\phi$'s:
\bq
\label{lower_terms}
 l_j
 = 
 \frac{t}{4 \lambda}
 \left[
 \alpha^2 \left( \phi_{x_j-ab_0} + \phi_{x_j+ab_0} \right)
 + 2 \left( D + \frac{m^2}{2} -1 - \alpha^2 \right) \phi_{x_j}
 - \sum\limits_{k=1}^{D-1}  \left( \phi_{x_j-ab_k} + \phi_{x_j+ab_k} \right)
 \right].
\eq
We consider the ideal
\bq
 J
 & = &
 \left\langle f_1, \dots, f_N \right\rangle
\eq
in the polynomial ring ${\mathbb C}[\phi_{x_1},\dots,\phi_{x_N}]$.
It is not too difficult to see that
the set $\{f_1,\dots,f_N\}$ is a reduced Gr\"obner basis for $J$ with respect
to the graded reverse lexicographic order (or the graded lexicographic order).
We are in the lucky situation that we get a reduced Gr\"obner basis for free.
By polynomial division we may write
\bq
 \hat{\Phi} & = & r + \sum\limits_{j=1}^N p_j f_j,
\eq
where the remainder $r$ is at most of degree two in any variable $\phi_x$.
This representation is unique and involves only commutative algebra.
Integration-by-parts allows us to replace the polynomial $\hat{\Phi}$ by (see eq.~(\ref{equivalence_relation}))
\bq
 \hat{\Phi}'
 & = &
 r 
 - \frac{i \alpha}{4 \lambda} \sum\limits_{j=1}^N \partial_j p_j,
\eq
where we used the short-hand notation $\partial_j = \frac{\partial}{\partial_{\phi_{x_j}}}$.
By construction, the polynomial $r$ is already reduced (i.e. the polynomial $r$ is a linear combination of the $\hat{e}_j$'s).
On the second term we may use recursion.
This algorithm can be implemented efficiently in {\tt FORM} \cite{FORM,Ruijl:2017dtg} or commercial computer algebra systems. 

\end{appendix}

{\footnotesize
\bibliography{/home/stefanw/notes/biblio}
\bibliographystyle{/home/stefanw/latex-style/h-physrev5}
}

\end{document}